\documentclass[aps,pra,showpacs,twocolumn,superscriptaddress]{revtex4-1}
\usepackage{bm,color,amsmath,amssymb,mathrsfs,latexsym,graphicx,psfrag,float}

\newcommand{\opt}[1]

\usepackage{comment}
\usepackage{enumerate}
\usepackage{txfonts}
\usepackage{bbm}
\usepackage{ifthen}
\usepackage{cleveref}
\usepackage{dsfont}
\usepackage{tabularx}
\newcolumntype{L}[1]{>{\raggedright\arraybackslash}p{#1}} 
\newcolumntype{C}[1]{>{\centering\arraybackslash}p{#1}} 
\newcolumntype{R}[1]{>{\raggedleft\arraybackslash}p{#1}} 

\newcommand{\eq}[2]{\begin{eqnarray}\label{#1}& #2& \end{eqnarray}}

\newcommand{\cre}[2]{#1^{\dagger}_{#2}}
\newcommand{\crea}[2]{#1^{*}_{#2}}

\newcommand{\ann}[2]{#1^{\phantom{\dagger}}_{#2}}
\newcommand{\sub}[1]{_{\mbox{\tiny #1}}}
\newcommand{\intf}[0]{\int\frac{d\omega}{2\pi}}

\usepackage{color}

\begin{document}

\title{Dicke Quantum Spin and Photon Glass in Optical Cavities: \\
Non-equilibrium theory and experimental signatures}

\author{Michael Buchhold}
\affiliation{Institut f\"ur Theoretische Physik, Leopold-Franzens Universit\"at Innsbruck, A-6020 Innsbruck, Austria}
\author{Philipp Strack}
\affiliation{Department of Physics, Harvard University, Cambridge MA 02138}
\author{Subir Sachdev}
\affiliation{Department of Physics, Harvard University, Cambridge MA 02138}
\author{Sebastian Diehl}
\affiliation{Institut f\"ur Theoretische Physik, Leopold-Franzens Universit\"at Innsbruck, A-6020 Innsbruck, Austria}
\affiliation{Institute for Quantum Optics and Quantum Information of the Austrian Academy of Sciences, A-6020 Innsbruck, Austria}

\begin{abstract}

In the context of ultracold atoms in multimode optical cavities, the appearance of a quantum-critical glass phase of atomic spins has been predicted recently. Due to the long-range nature of the cavity-mediated interactions, but also the presence of a driving laser and dissipative processes such as cavity photon loss, the quantum optical realization of glassy physics has no analog in condensed matter, and could evolve into a ``cavity glass microscope'' for frustrated quantum systems out-of-equilibrium. 
Here we develop the non-equilibrium theory of the multimode Dicke model with quenched disorder and Markovian dissipation.
Using a unified Keldysh path integral approach, we show that the defining features of a low temperature glass, representing a critical phase of matter with algebraically decaying temporal correlation functions, are seen to be robust against the presence of dissipation due to cavity loss. The universality class however is modified due to the Markovian bath. The presence of strong disorder leads to an enhanced equilibration of atomic and photonic degrees of freedom, including the emergence of a common low-frequency effective temperature. The imprint of the atomic spin glass physics onto a ``photon glass'' makes it possible to detect the glass state by standard experimental techniques of quantum optics. We provide an unambiguous characterization of the superradiant and glassy phases in terms of fluorescence spectroscopy, homodyne detection, and the temporal photon correlation function $g^{(2)}(\tau)$.

\end{abstract}


\maketitle
\section{Introduction}

An emerging theme in the research on strongly correlated ultracold atoms is the creation of 
quantum soft matter phases ranging from nematics and smectics \cite{Gopalakrishnan2009,Gopalakrishnan2010}, 
liquid crystals \cite{lechner12}, granular materials \cite{igor11,pohl11,peyronel12}, friction phenomena in nonlinear lattices \cite{prutti11,vitelli13}, to glasses \cite{Gopalakrishnan2011,Strack2011,Strack2012,PhysRevLett.109.020403,poletti12}. Realizing 
glasses with strongly interacting light-matter systems bears the promise to study some of the most celebrated achievements in statistical mechanics from a new vantage point. The Parisi solution of mean-field spin glasses \cite{BinderYoung}, for example, continues to trigger research more than three decades after its discovery in the early 1980's, and may have implications for information storage \cite{Amit1985} and ``frustrated'' optimization algorithms \cite{bapst13}. The latter is related to the inability of a glass to find its ground state; a feature that makes it inherently non-equilibrium.

Historically, quantum effects in soft matter and glasses have not played a prominent role because most soft materials are too large, too heavy and/or too hot and therefore way outside the quantum regime. Spin and charge glass features have however been invoked in some electronic quantum materials \cite{BinderYoung,miranda05}, mainly due to RKKY-type interactions or randomly distributed impurities providing a random potential for the electrons. However, here the glassy mechanisms occur often in combination with other more dominant (Coulomb) interactions, and it is hard to pin down which effects are truly due to glassiness. Note that the somewhat simpler Bose glass of the Bose Hubbard model \cite{mpafisher89} (see \cite{PhysRevLett.110.075304} for a possible realization in optical cavities), while in the quantum regime, occurs because of a short-range random potential, and does not generically exhibit some of the hallmark phenomena of frustrated glasses, such as many metastable states, aging, or replica-symmetry breaking.

It would clearly be desirable to have a tunable realization of genuinely frustrated (quantum) glasses in the laboratory. 
Recent work on ultracold atoms in optical cavities \cite{Gopalakrishnan2011,Strack2011,Strack2012,PhysRevLett.110.075304} suggests that it may be possible to create spin- and charge glasses in these systems, which arise because of frustrated couplings of the atomic ``qubits'' to the dynamical potential of multiple cavity modes. It is appealing to these systems that the photons escaping the cavity can be used for in-situ detection of the atom dynamics (``cavity glass microscope''), and that the interaction mediated by cavity photons is long-ranged. The latter makes the theoretical glass models more tractable and should allow for a realistic comparison of experiment and theory.

The phases of matter achievable with cavity quantum electrodynamics (QED) systems settle into {\it non-equilibrium steady states} typically balancing a laser drive with 
dissipation channels such as cavity photon loss and atomic spontaneous emission. The notion of temperature is, a priori, not well defined.
A line of recent research on the self-organization transition of bosonic atoms in a single-mode optical cavity (experimentally realized with a thermal gas of Cesium \cite{black2003} and with a Bose-Einstein condensate of Rubidium \cite{Brennecke07, baumann2010}), has established the basic properties of the non-equilibrium phase transition into the self-organized, superradiant phase \cite{domokos2002, Mekhov07,keeling2010,nagy2010,nagy11,oztop2012,bhaseen2012,Dalla2012,Mottl12, RitschReview}. In particular it was shown that, upon approximating the atom dynamics by a single collective spin of length $N/2$ and taking the 
atom number $N$ large, the dynamics can be described by classical equations of motion \cite{keeling2010,bhaseen2012}, and that the phase transition becomes thermal due to the drive and dissipation \cite{Dalla2012}. 

In this paper, we underpin our previous proposal \cite{Strack2011}, and show that quenched disorder from multiple cavity modes, leads to qualitatively different non-equilibrium steady states with quantum glassy properties. We develop a comprehensive non-equilibrium theory for many-body multimode cavity QED with quenched disorder and Markovian dissipation. We pay special attention to the quantum optical specifics of the pumped realization of effective spin model \cite{Dimer2007}, the laser drive and the finite photon lifetimes of cavity QED. Using a field theoretic Keldysh formalism adapted to quantum optics, we compute several observables of the glass and superradiant phases, which are accessible in experiments with current technology. Our key results are summarized in the following section.

The remainder of the paper is then organized as follows. 
In Sec.~\ref{sec2} we discuss the multimode open Dicke model in the simultaneous presence of quenched disorder and Markovian dissipation. Disorder and dissipative baths are contrasted more rigorously in App. \ref{sec:MvsQ}. We switch to a unified description of both these aspects in Sec.~\ref{sec:pathintegral} in the framework of a Keldysh path integral formulation, and specify the formal solution of the problem in the thermodynamic limit in terms of the partition sum, which allows to extract all atomic and photonic correlation and response functions of interest. This solution is evaluated in Sec.~\ref{sec:Results}, with some details in App. \ref{sec:XDist}. This comprises the calculation of the phase diagram in the presence of cavity loss, as well as the discussion of correlation and response functions for both atomic and photonic degrees of freedom, allowing us to uniquely characterize the simultaneous spin and photon glass phase from the theoretical perspective.  We then discuss the consequences of these theoretical findings to concrete experimental observables in cavity QED experiments in Sec.~\ref{sec:Exp}. The combination of correlation and response measurements allows for a complete characterization of the phase diagram and in particular of the glass phase.

The relation between Keldysh path integral and quantum optics observables is elaborated on further in App. \ref{sec:LinResp}.

\section{ Key Results} 
\label{sec:keyresults}

\subsection{Non-equilibrium steady state phase diagram}

\begin{figure}[t!]
\includegraphics[width=0.95\linewidth]{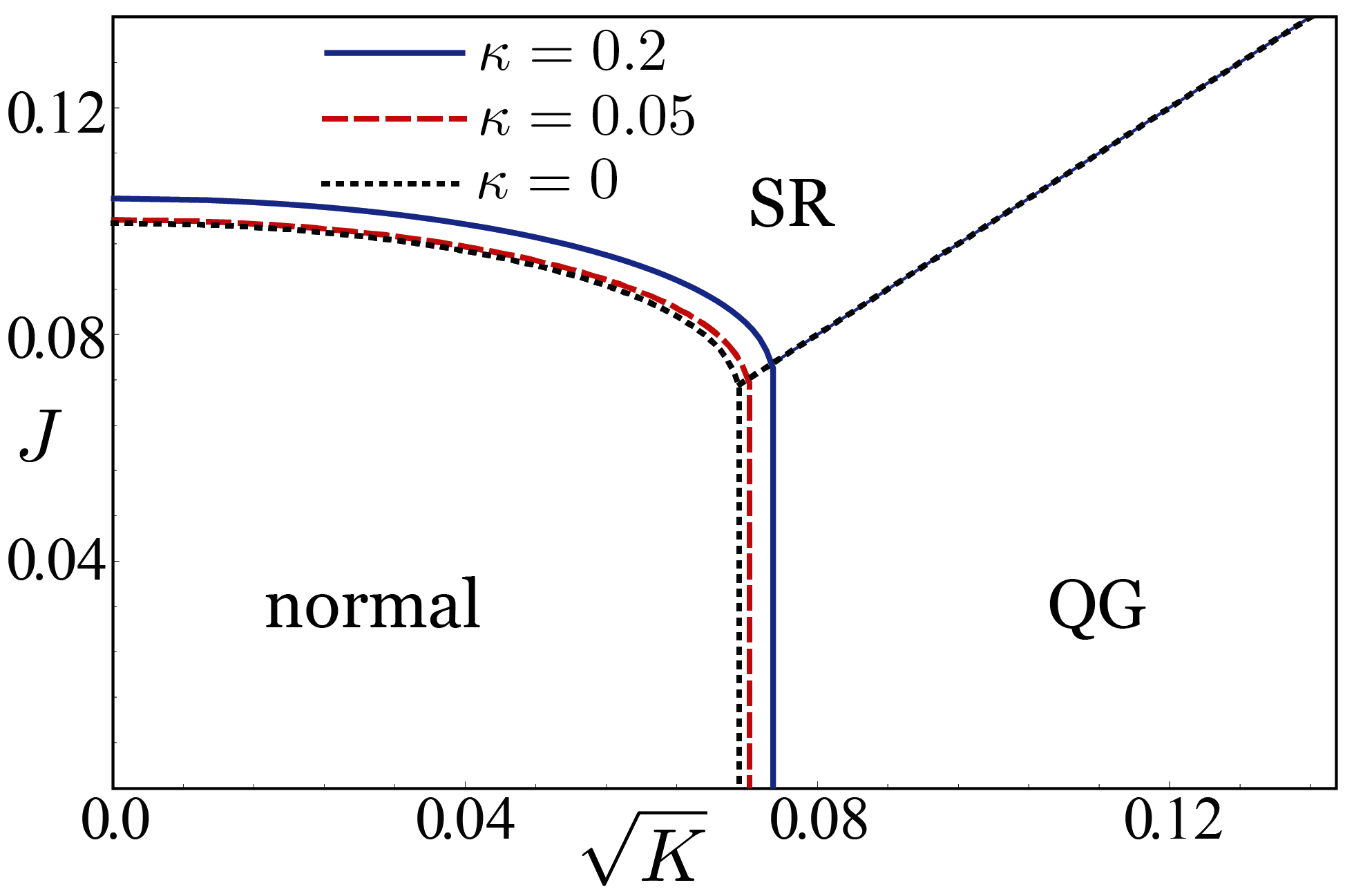}
    \caption
    {\label{fig:PhaseDiag}
    (Color online) {\it Non-equilibrium steady state phase diagram} of the open multimode Dicke model, as a function of averaged atom-photon coupling $J$ ($y$-axis) and disorder variance $K$ ($x$-axis) and  for parameters ${\omega_0=1}$ (cavity detuning) and ${\omega_z=0.5}$ (effective atom detuning) for different photon decay rates $\kappa$. QG is the quantum spin and photon glass, SR the superradiant phase. The $T=0$ equilibrium phase diagram of Ref. \cite{Strack2011} is recovered as $\kappa \to 0$. The SR-QG transition is not affected by $\kappa$.}
\end{figure}

\emph{The shape of the phase diagram for the steady state predicted in \cite{Strack2011} , with the presence of a normal, a superradiant, and a glass phase is robust in the presence of Markovian dissipation, cf. Fig. \ref{fig:PhaseDiag}.}
As to the phase diagram, the open nature of the problem only leads to quantitative modifications. In particular, the characteristic feature of a glass representing a critical phase of matter persists. The presence of photon decay 
overdamps the spin spectrum and changes the universality class of the glass phase, which we now discuss.

\subsection{Dissipative spectral properties and universality class}\label{sec:disssp}
\begin{figure}[t!]
\includegraphics[width=0.97\linewidth]{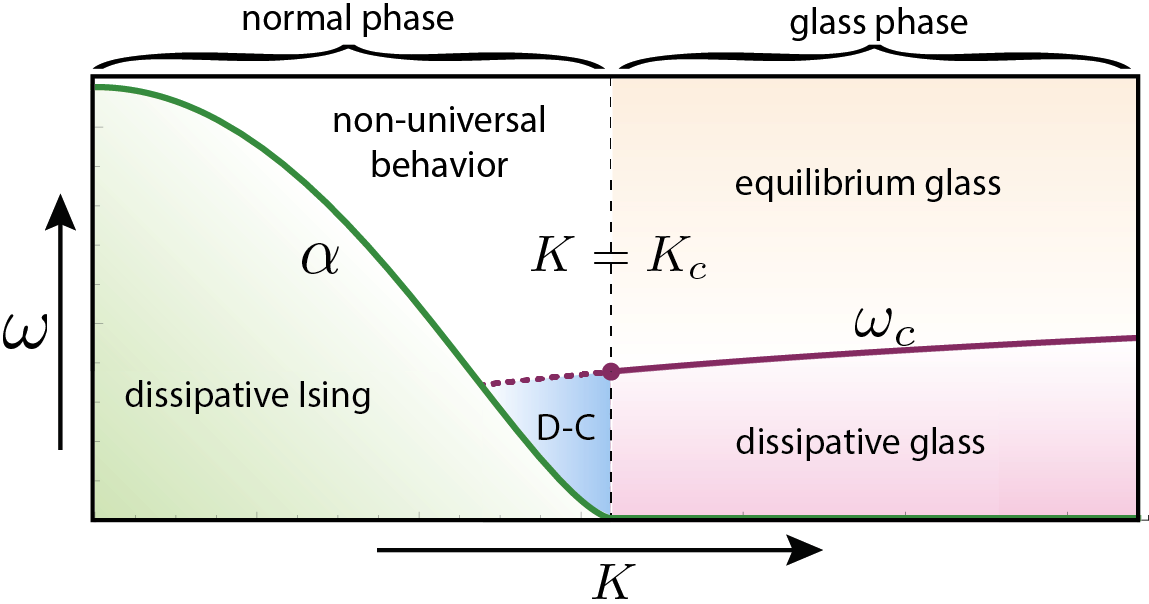}
    \caption
    {\label{fig:FRIllu}
    (Color online)  {\it Illustration of the dissipative spectral properties and universality class.} 
As a function of probe frequency $\omega$ ($y$-axis) and the disorder variance $K$ ($x$-axis), we illustrate the different regimes in the phase diagram. In the normal phase, for frequencies $\omega<\alpha$ the system is represented by a dissipative Ising model, described by Eq.~\eqref{KeyEq3}, while for frequencies $\omega>\alpha,\omega_c$ it is described by non-universal behavior of a disordered spin fluid. In the glass phase ($K>K_c$), there exist two qualitatitvely distinct frequency regimes, separated by the crossover scale $\omega_c$, cf. Eq.~\eqref{KeyEq1}. At the lowest frequencies, $\omega <\omega_c$ the system is described by the universality class of dissipative spin glasses. For $\omega>\omega_c$, we find that the system behaves quantitatively as an equilibrium spin glass. For $\alpha<\omega_c$ and $K<K_c$, there exists a dissipative crossover region (D-C in the figure), which is a precursor of the dissipative spin glass. It shows dissipative Ising behavior for smallest frequencies and resembles the dissipative glass for frequencies $\omega_c>\omega>\alpha$. 
}
\end{figure}

\begin{figure}[t!]
\includegraphics[width=0.97\linewidth]{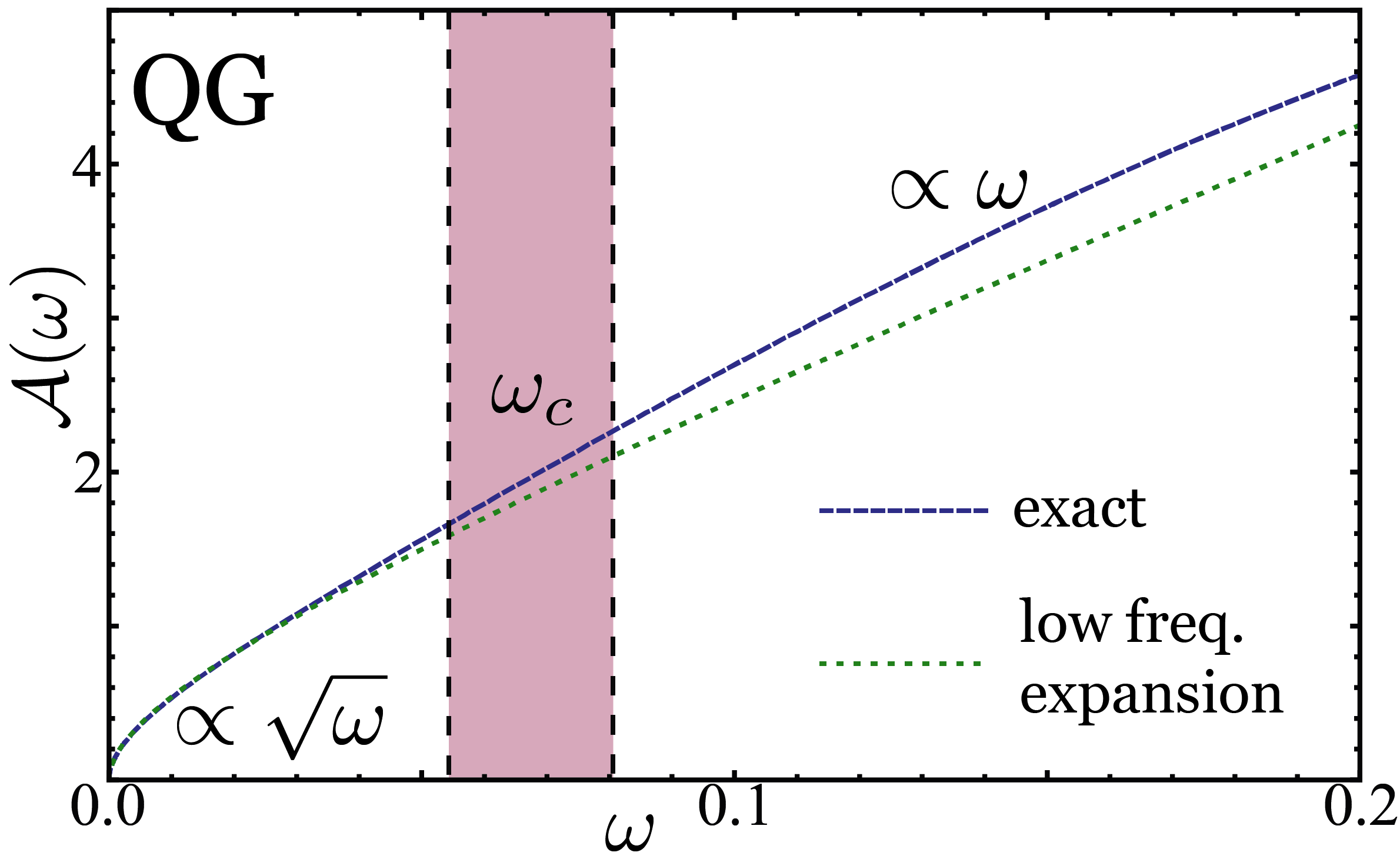}
    \caption
    {\label{fig:QGSpec}
    (Color online) {\it Dissipative spectral properties and universality class} of the single-atom spectral density ${\mathcal{A}(\omega)}$ (response signal of RF spectroscopy) in the quantum glass phase for parameters ${K=0.01, J=0.1, \omega_z=2, \kappa=0.1,}$ $\omega_0=0.7$. For frequencies ${\omega<\omega_c}$ below the crossover scale, the spectral density is overdamped and proportional to $\sqrt{\omega}$.
  For intermediate frequencies ${\omega>\omega_c}$, $\mathcal{A}$ is linear in the frequency, as for the non-dissipative case \cite{Strack2011}, which is recovered in the limit ${\kappa\rightarrow0}$.}
\end{figure}

\emph{Within the glass phase, we identify a crossover scale $\omega_c\sim \kappa$ proportional to the cavity decay rate $\kappa$, above which the spectral properties of a zero temperature quantum spin glass are reproduced. Although the finite cavity decay $\kappa$ introduces a finite scale ``above the quantum critical point of the closed, equilibrium system'', $\kappa$ acts very differently from a finite temperature. In particular, below $\omega_c$, the spectral properties are modified due to the breaking of time reversal symmetry by the Markovian bath, while remaining critical. Due to the low frequency modification, the quantum spin glass in optical cavities formally belongs to the dynamical
universality class of dissipative quantum glasses, such as glasses coupled to equilibrium ohmic baths \cite{Cugliandolo1999,Cugliandolo2002,Cugliandolo2004} or  metallic spin glasses \cite{Read1995,Strack2012}.}

\emph{Spectral properties} --  The role of the crossover scale between equilibrium and dissipative spin glass is further illustrated in Fig. \ref{fig:FRIllu}. It is given by
\begin{eqnarray}
\omega_\text{c} = 2\kappa\left(1+\frac{\omega_0^2}{\omega_0^2+\kappa^2}+\frac{\left(\omega_0^2+\kappa^2\right)^2}{\sqrt{K}\omega_z^2}\right)^{-1}.\label{KeyEq1}
\end{eqnarray}
 The resulting modifications below this scale, compared to a more conventional equilibrium glass are due to the Markovian bath, introducing damping. In the normal and superradiant phases, this allows for the following form of the frequency resolved linearized Langevin equation for the atomic Ising variables,
\begin{eqnarray}\label{KeyEq3}
&\frac{1}{Z}\left(\omega^2+ i\gamma\omega+\alpha^2\right) x(\omega) = \xi (\omega),&
\end{eqnarray}
modelling the atoms as an effective damped harmonic oscillator, with finite life-time ${\tau=\frac{1}{\gamma}<\infty}$ and $\alpha$ the effective oscillator frequency, with the physical meaning of the gap of the atomic excitations in our case. The noise has zero mean and $\langle \xi (t') \xi(t)\rangle = \frac{2\gamma}{Z}T_\text{eff} \delta (t' - t)$. 

At the glass transition, $Z$ and $\alpha$ scale to zero simultaneously and the frequency dependence becomes gapless and non-analytic. In the entire glass phase, the effective atomic low frequency dynamics is then governed by the form
\eq{KeyEqEx1}{
\frac{1}{\bar{Z}}\sqrt{\omega^2+\bar{\gamma}|\omega|} \,\, x(\omega) = \xi (\omega) ,}
which obviously cannot be viewed as a simple damped oscillator any more.  
The broken time reversal symmetry manifests itself in $\gamma, {\bar{\gamma}>0}$, thus modifying the scaling for ${\omega \to 0}$. The crossover between these different regimes is clearly visible in Fig. \ref{fig:QGSpec}.

\emph{Universality class} -- The qualitative modification of the low-frequency dynamics below the crossover scale $\omega_\text{c}$ implies a modification of the equilibrium quantum spin glass universality class. The open system Dicke superradiance phase transition, where the $Z_2$ symmetry of the Dicke model is broken spontaneously due to a finite photon condensate, is enclosed by a finite parameter regime in which the dynamics is purely dissipative, or overdamped (see e.g. \cite{Dalla2012}). Together with the generation of a low frequency effective temperature (LET), for this reason the single mode Dicke phase transition can be classified within the scheme of Hohenberg and Halperin \cite{Hohenberg} in terms of the purely relaxational Model A, thereby sharing aspects of an equilibrium dynamical phase transition. This situation is different for the open system glass transition: Here, irreversible dissipative and reversible coherent dynamics rival each other at the glass transition down to the lowest frequencies. In particular, the dissipative dynamics fades out \emph{faster} than the coherent dynamics as witnessed by larger critical exponents, and there is no regime in the vicinity of the critical point where either dissipative or coherent dynamics would vanish completely. This behavior is demonstrated in the inset of Fig. \ref{fig:Fluorescence}.

We note that, while these findings are unconventional from the viewpoint of equilibrium quantum glasses, they are not uniquely tied to the presence of the driven, Markovian non-equilibrium bath. In fact, such behavior is also present in the case of a system-bath setting in global thermodynamic equilibrium, where the presence of the bath variables modifies the spectral properties of the spins \cite{Cugliandolo2002, Cugliandolo1999, Strack2012}. Both physical contexts share in common the time reversal breaking of the subsystem obtained after elimination of the bath modes and may be seen to belong to the same universality class.

\subsection{Atom-photon thermalization into quantum-critical regime}

\begin{figure}[t!]
\includegraphics[width=1.\linewidth]{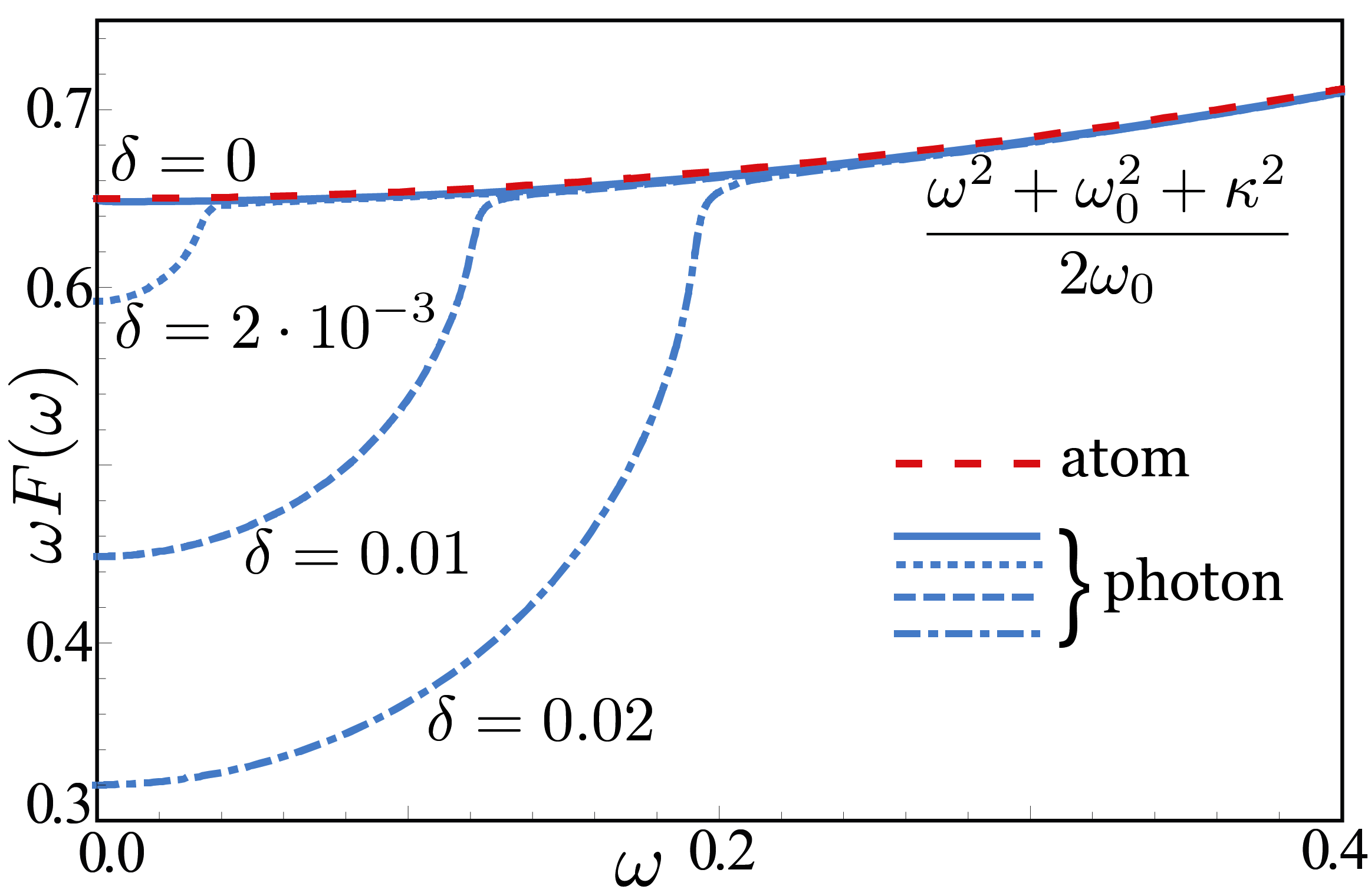}
    \caption
    {\label{fig:Thermalization}
    (Color online) {\it Thermalization into quantum-critical regime} of the atomic (red, dashed line) and photonic (blue lines) distribution functions 
    $F(\omega)$ when approaching the glass transition at a critical disorder variance $K_c$ for $\omega_0=1.3, \omega_z=0.5, \kappa=0.01, K_c=0.01$, $J=0.1$ and varying parameter $\delta=K_c-K$. For larger values of $J$, i.e. larger distance from the glass transition, the low frequency effective temperature (LET) $ 2 T_\text{eff} = \lim_{\omega \to 0} \omega F(\omega)$ of the photons is much lower than the LET of the atoms and the frequency interval for which atoms and photons are not equilibrated is larger. When the glass transition is approached, atoms and photons attain the same LET.}
\end{figure}

\emph{As in the driven open Dicke model, the statistical properties of atoms and photons are governed by effective temperatures at low frequencies. The effective temperature differs in general for the two subsystems. Approaching the glass transition, these effective temperatures are found to merge. The finite cavity decay enables this mechanism but $\kappa$ does not directly play the role of effective temperature. This mechanism pushes the hybrid system of atoms and photons in the glass phase into a 
quantum-critical regime described by a global effective temperature for a range of frequencies. This quantum critical regime retains signatures of the underlying quantum critical point.}

The Markovian bath not only affects the spectral properties, but also governs the statistical properties of the system. The main statistical effect is the generation of a LET for the atomic degrees of freedom, for which we find 
\begin{eqnarray}
T_\text{eff} = \frac{\omega_0^2+\kappa^2}{4\omega_0}\label{KeyEq2}
\end{eqnarray}throughout the entire phase diagram, and taking the same value as in the single-mode case (in the absence of spontaneous emission for the atoms). This thermalization of the atoms happens despite the microscopic driven-dissipative nature of the dynamics, and has been observed in a variety of driven open systems theoretically \cite{mitra06,Diehl10,dallatorre10, oztop2012, dalla_ehud12, Dalla13, wouters2006, Sieberer2013} and experimentally \cite{weitz10}.  Below this scale, the occupation properties are governed by an effective classical thermal distribution $2T_\text{eff}/\omega$, while above it the physics is dominated by non-equilibrium effects \cite{Dalla2012}. For cavity decay $\kappa\ll \omega_0$, the crossover scale obeys $\omega_c\ll T\sub{eff}$. As a consequence, in an extended regime of frequencies between $\omega_c$ and $T_{\text{eff}}$, the correlations describe a finite temperature equilibrium spin glass.

In the single-mode open Dicke model, the photon degrees of freedom are also governed by an effective temperature, which however differs from the one for atoms \cite{Dalla2012}, indicating the absence of equilibration between atoms and photons even at low frequencies. The increase of the disorder variance leads to an adjustment of these two effective temperatures, cf. Fig.  \ref{fig:Thermalization}. At the glass transition, and within the entire glass phase, the thermalization of the subsystems is complete, with common effective temperature given in Eq. (\ref{KeyEq2}). This effect  can be understood qualitatively as a consequence of the disorder induced long ranged interactions, cf. Sec. \eqref{2Eq35}. These allow to redistribute energy and enable equilibration.

We emphasize that the notion of thermalization here refers to the expression of a $1/\omega$ divergence for the system's distribution function, as well as the adjustment of the coefficients for atoms and photons. This provides an understanding for distinct scaling properties of correlations (where the distribution function enters) vs. responses (which do not depend on the statistical distribution), which can be addressed separately in different experiments (see below). Crucially, this notion of ``thermalization'' does not mean that the characteristics features of the glass state are washed out or overwritten.

 \subsection{Emergent photon glass phase}

\begin{figure}[t]
\vspace{1mm}
\includegraphics[width=1\linewidth]{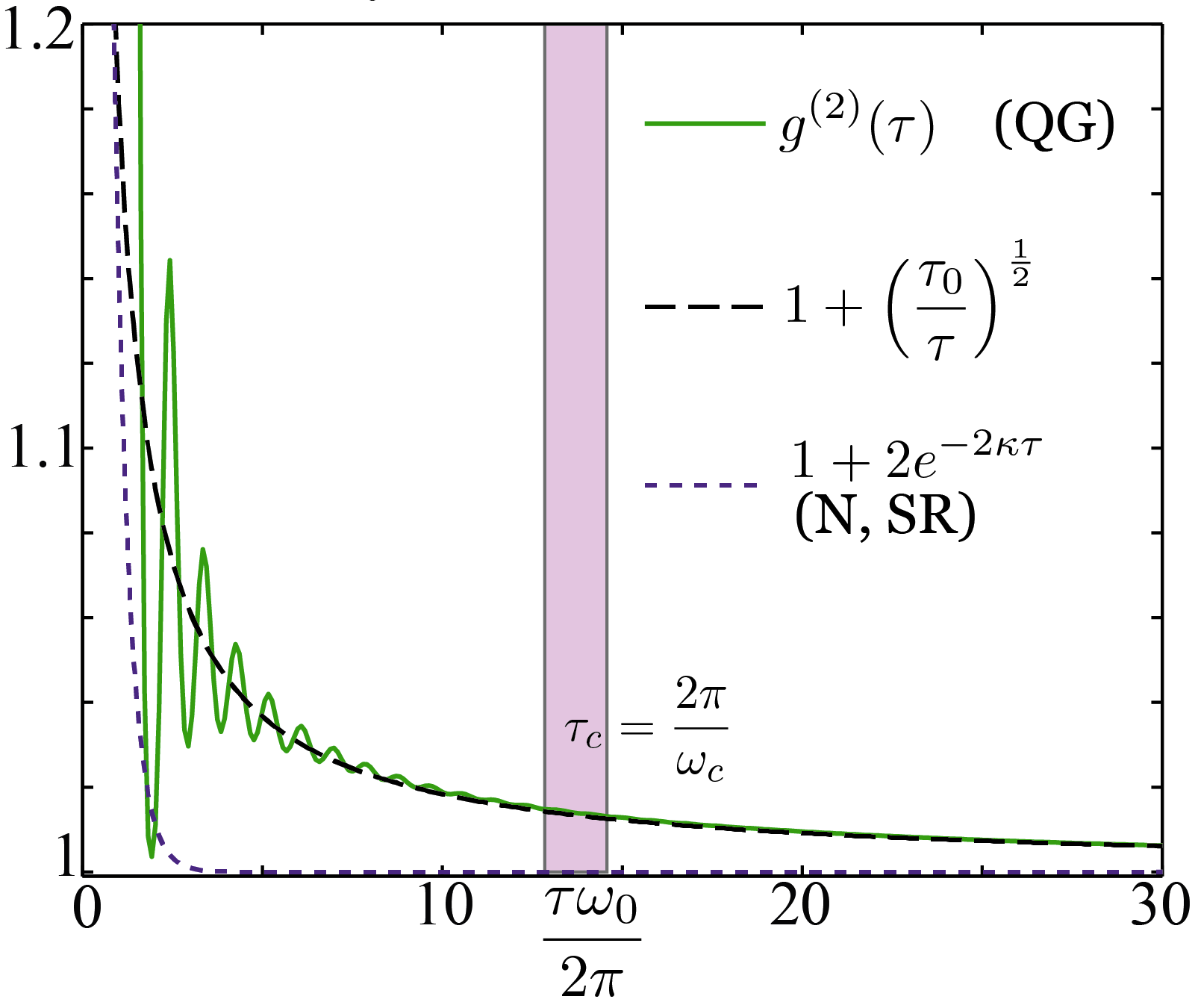}
    \caption
    {\label{fig:g2}
    (Color online) {\it Emergent photon glass phase} with algebraically decaying  photon correlation function $g^{(2)}(\tau)$ at long times, for parameters $\omega_0=1, \kappa=0.4, \omega_z=6, J=0.4, K=0.16$. The time-scale for which algebraic decay sets in is determined by the inverse crossover frequency $\omega_c$, given by Eq.~\eqref{4Eq7}. For comparison, we have also plotted the envelope of the exponential decay of the correlation function in the normal and superradiant phase.
The short time behavior of the correlation function is non-universal and not shown in the figure, however, $g^{(2)}(0)=3$ due to the effective thermal distribution for low frequencies. The parameter $\tau_0=O(\frac{1}{\omega_0})$ was determined numerically.}
\end{figure}

\emph{The strong light-matter coupling results in a complete imprint of the glass features of the atomic degrees of freedom 
onto the photons in the cavity. We refer to the resulting state of light as a photon glass highlighting the connection of multimode cavity QED to random lasing media \cite{Angelani2006,Andreasen2011}.}

The photon glass is characterized by a photonic Edwards-Anderson order parameter signaling infinitely long memory in certain temporal two-point correlation function. This implies that a macroscopic number of photons is permanently present in the cavity (extensive scaling with the system size), which are however not occupying a single mode coherently, but rather a continuum of modes.
The presence of a continuum of modes at low freqeuency is underpinned by the slow algebraic decay of the system's correlation functions 
as shown for the photon correlation function in Fig.~\ref{fig:g2}. This is a consequence of the disorder-induced degeneracies. 
$g^{(2)}(\tau)$ is accessible by detecting the photons that escape the cavity.

\subsection{Cavity glass microscope}

\begin{figure}[t!]
\includegraphics[width=1\linewidth]{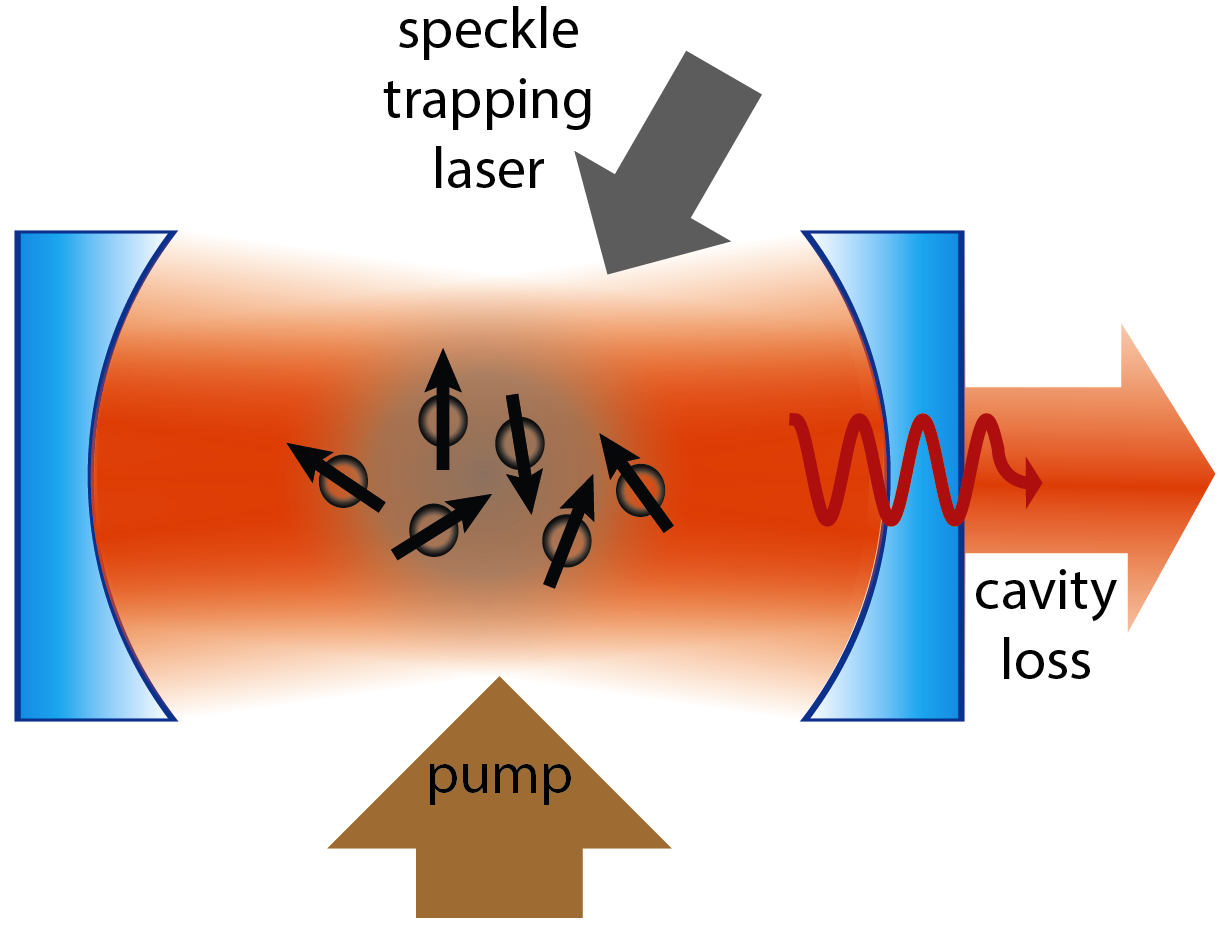}
    \caption
    {\label{fig:cavity}
    (Color online) {\it Cavity glass microscope} set-up: Atoms are placed in a multimode cavity subject to a transversal laser drive with pump frequency $\omega_p$. The atoms are fixed at random positions by an external speckle trapping potential over regions inside the cavity, wherein mode functions $g({\bf k}_i, {\bf x}_l)$ randomly change sign as a function of the atomic positions, in order to provide frustration, as well as vary in magnitude. The more cavity modes, the better, and in particular the regime where the ratio of the number of cavity modes ($M$) over the number of atoms ($N$), $\alpha = M/N$ is kept sizable is a promising regime for glassy behavior \cite{Amit1985, Gopalakrishnan2011}. Photons leaking from the cavity with rate $\kappa$ give rise to additional dissipative dynamics and allow for output detection measurements. }
\end{figure}

\begin{figure}[t!]
\includegraphics[width=1\linewidth]{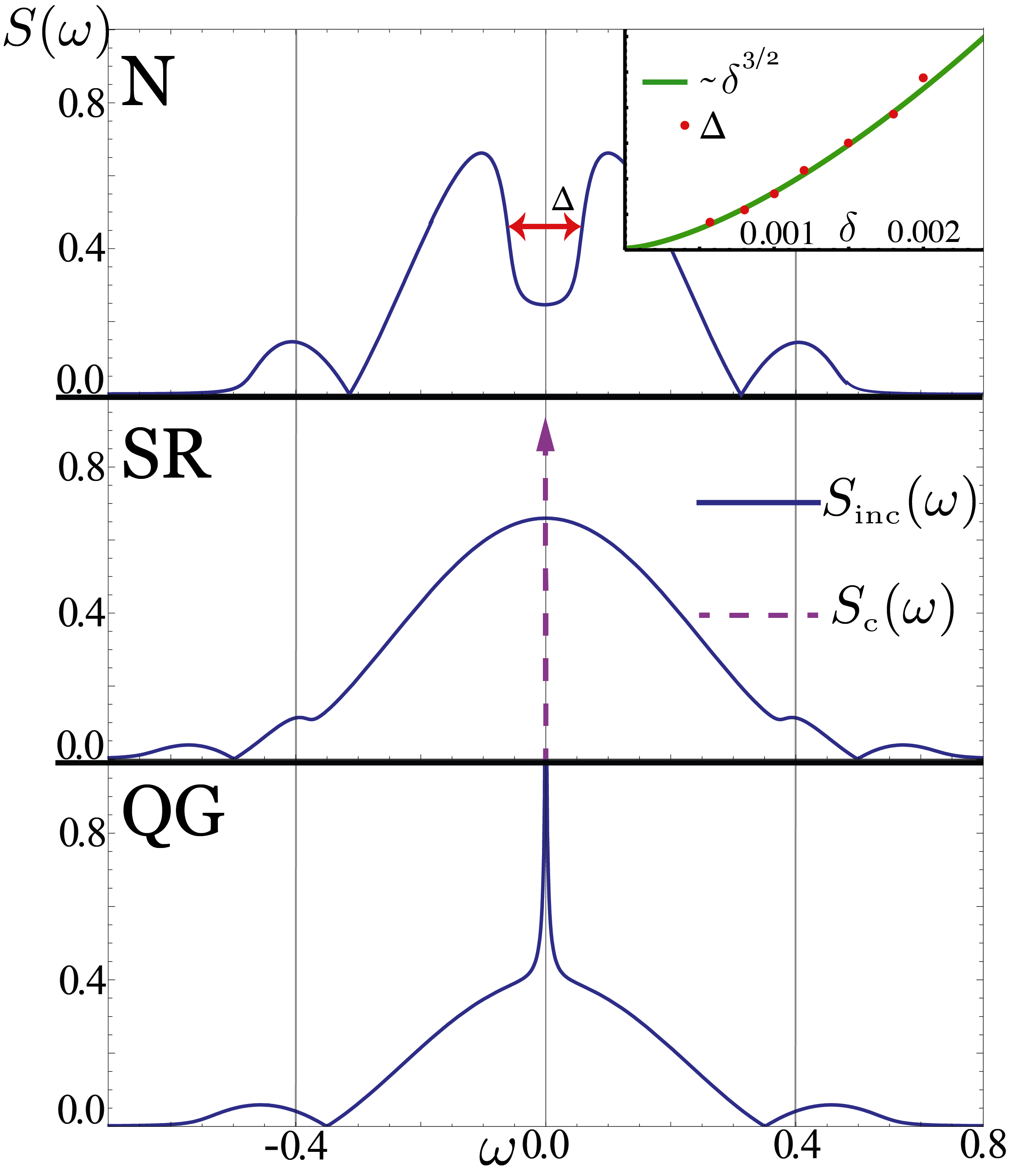}
    \caption
    {\label{fig:Fluorescence}
    (Color online) {\it Cavity glass microscope} output of a typical fluorescence spectrum $S(\omega)$ (not normalized), decomposed in coherent $S\sub{c}$ and incoherent part $S\sub{inc}$ for the three distinct phases in the multimode Dicke model. The parameters $J, K$ are varied, while $\omega_0=1, \kappa=0.1, \omega_z=0.5$ are kept fixed for each panel.\\ \emph{Normal phase}, $(J,K)=(0.13, 0.008)$. Central and outer doublets are visible but broadened by the disorder, only the incoherent contribution is non-zero.\\ 
\emph{Superradiant phase}, $(J,K)=(0.4, 0.008)$. The central doublets have merged due to the presence of a critical mode at $\omega=0$. At zero frequency there is a coherent $\delta$-contribution indicated by the arrow (dashed).\\ \emph{Glass phase}, $(J,K)=(0.13, 0.017)$. There is a characteristic $\frac{1}{\sqrt{\omega}}$ divergence for small $\omega<\omega_c$ due to the non-classical critical modes at zero frequency. The peak at $\omega=0$ is incoherent and can therefore easily be discriminated from the coherent peak in the middle panel.\\
\emph{Scaling of correlations}. The inset in the upper panel shows the behavior of the peak distance of $S(\omega )$ in the normal phase when approaching the glass phase. The two peaks approach each other and merge at the glass transition. The distance follows the dominant coherent exponent $\alpha_{\delta}\propto\delta^{\frac{3}{2}}$, cf. Sec. \ref{sec:disssp}.  }
\end{figure}

\emph{ The cavity set-up of Fig.~\ref{fig:cavity} should allow for unprecedented access to the strongly coupled light-matter phase 
with disorder. Adapting the input-output formalism of quantum optics \cite{Collett1984, Gardiner1985} to the Keldysh path integral, 
we provide a comprehensive experimental characterization of the various phases in terms of the cavity output spectrum and 
the photon correlations $g^{(2)}(\tau)$ in the real time domain. }

This continues and completes a program started in \cite{Dalla2012} of setting up a  translation table between the language and observables of quantum optics, and the Keldysh path integral approach. The frequency and time resolved \emph{correlations} can be determined via fluorescence spectroscopy, cf. Fig.~\ref{fig:Fluorescence}, and the measurement of $g^{(2)}(\tau)$ follows time-resolved detection of cavity output, cf. Fig. \ref{fig:g2}.
The fluorescence spectrum shows a characteristic $\frac{1}{\sqrt{\omega}}$ divergence for small frequencies $\omega<\omega_c$. 
This indicates a macroscopic but incoherent occupation of the cavity as anticipated above: The glass state is not characterized by a single-particle order parameter where a single quantum state is macroscopically occupied, and which would result in (temporal) long range order such as a superradiant condensate. Rather it is characterized by a strong and infrared divergent occupation of a continuum of modes, giving rise to temporal quasi-long range order. This phenomenology is reminiscent of a Kosterlitz-Thouless critical phase realized e.g. in low temperature weakly interacting Bose gases, with the difference that spatial correlations are replaced by temporal correlations.

Finally, the combined measurement of response and correlations enables the quantitative extraction of the effective temperature.

\section{multimode Open Dicke Model}\label{sec2}
In this section, we explain the model for fixed atoms in an open multimode cavity subject to a transversal laser drive shown in Fig.~\ref{fig:cavity}. 
We first write down the explicitly time-dependent Hamiltonian operator for a level scheme involving two Raman transitions proposed by Dimer 
{\it et al.} \cite{Dimer2007}. We then transform this Hamiltonian to a frame rotating with the pump frequency. This eliminates the explicit time dependencies in the Hamiltonian at the expense of violating detailed balance between the system and the electromagnetic bath surrounding the cavity. The bath becomes effectively Markovian in accordance with standard approximations of quantum optics. Finally, we eliminate the excited state dynamics to arrive at a multimode Dicke model with tunable couplings and frequencies.

\subsection{Hamiltonian operator}
\begin{figure}[t!]
\includegraphics[width=0.6\linewidth]{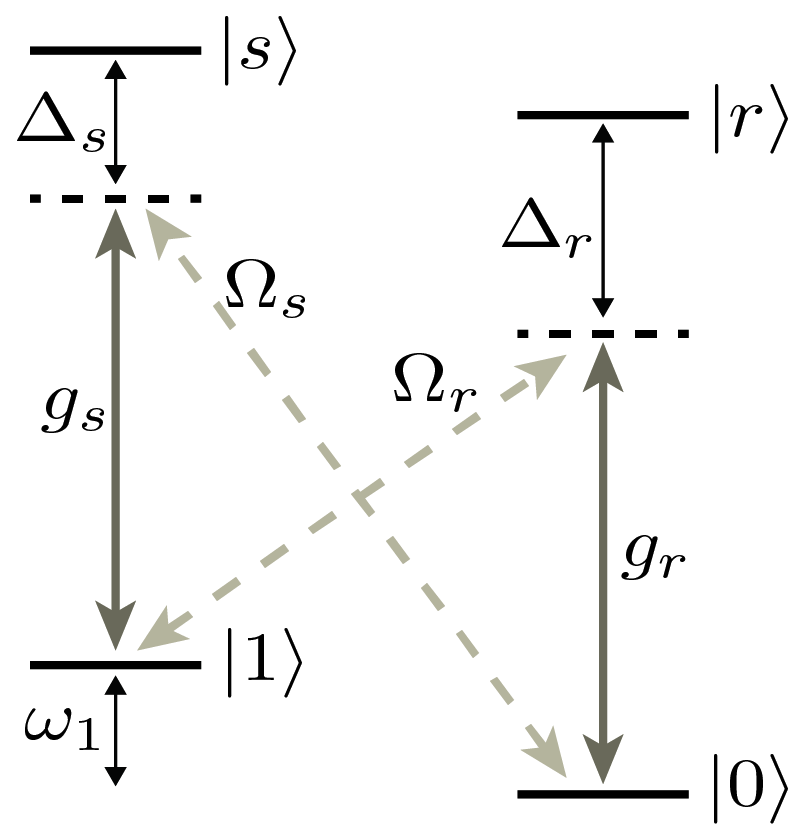}
    \caption
    {\label{fig:dimer}
    (Color online) Internal level scheme to generate tunable Dicke couplings between the ground state levels $|1\rangle$, $|0\rangle$ and the cavity. Adapted from Dimer {\it et al.} \cite{Dimer2007}.}
\end{figure}

The unitary time evolution of the atom-cavity system with the level scheme of Fig.~\ref{fig:dimer} follows the Hamiltonian
\begin{eqnarray}
&\hat{H}=\hat{H}_{\text{cav}}+\hat{H}_{\text{at}}+\hat{H}_{\text{int}}+\hat{H}(t)_{\text{pump}} ,&
\label{eq:H_full}
\end{eqnarray}
which we now explain one-by-one. The cavity contains $M$ photon modes with frequencies $\nu_i$
\begin{eqnarray}
&\hat{H}_{\text{cav}} = \sum_{i=1}^M \nu_i a^\dagger_i a_i ,&
\end{eqnarray}
which we later take to be in a relatively narrow frequency range $\nu_i\approx \nu_0$ such that the modes couple with comparable strengths to the detuned internal transition shown in Fig.~\ref{fig:dimer}. The atom dynamics with frequencies 
given relative to the lower ground state $|0\rangle$ is
\begin{align}
\hat{H}_{\text{at}}= \sum_{\ell=1}^N 
\omega_r | r_\ell \rangle \langle r_\ell |
+
\omega_s | s_\ell \rangle \langle s_\ell |
+
\omega_1 | 1_\ell \rangle \langle 1_\ell | .
\end{align}
The interaction between the atoms and cavity modes
\begin{align}
\hat{H}_{\text{int}}= 
\sum_{\ell=1}^N 
 \sum_{i=1}^M 
 \Big(
 g_r(\mathbf{k}_i, \mathbf{x}_\ell)
 | r_\ell \rangle \langle 0_\ell |
+
 g_s(\mathbf{k}_i, \mathbf{x}_\ell)
| s_\ell \rangle \langle 1_\ell |
\Big)
 \hat{a}_i
+
\text{H.c.}
\end{align}
involves a set of cavity mode functions $g(\mathbf{k}_i, \mathbf{x}_\ell)$ which depend on the wave vector of the cavity mode 
$\mathbf{k}_i$ and the position of the atom $\mathbf{x}_\ell$. The pump term 
\begin{eqnarray}
\hat{H}_{\text{pump}}(t)=\sum_{\ell = 1}^N
\Big(
&e^{-i\omega_{p, r} t}
\frac{ \Omega_r(\mathbf{k}_r, \mathbf{x}_\ell)}{2}
 | r_\ell \rangle \langle 1_\ell |
 \nonumber\\
 +&
 e^{-i\omega_{p,s} t }
 \frac{ \Omega_s(\mathbf{k}_s, \mathbf{x}_\ell)}{2}
 | s_\ell \rangle \langle 0_\ell |
 \Big)
+
\text{H.c.}
\label{eq:H_pump}
\end{eqnarray}
does not involve photon operators and induces coherent transitions between the excited and ground states as per Fig.~\ref{fig:dimer}. 
$\omega_p$ is the (optical) frequency of the pump laser. We assume the atoms to be homogeneously pumped from the side so that the mode function of the pump lasers are approximately constant $\Omega_{r,s}(\mathbf{k}_{r,s},\mathbf{x}_\ell)\approx \Omega_{r,s}$.

We now transform Eqs.~\eqref{eq:H_full}-\eqref{eq:H_pump} to a frame rotating with the (optical) frequency of the pump laser, mainly to eliminate the explicit time dependence from the pump term Eq.~\eqref{eq:H_pump}~\cite{Dimer2007}. The unitary transformation operator is 
$\hat{U}(t)=\exp (-i\hat{H}_0 t)$ with 
$\hat{H}_0 =\left( \omega_{p,s}-\omega_{1}'\right)\sum_{i=1}^M a^\dagger_i a_i 
+ 
\sum_{\ell =1}^N \Big\{ \left(\omega_{p,r}+\omega_1'\right)| r_\ell \rangle \langle r_\ell | + \omega_{p,s} | s_\ell \rangle \langle s_\ell |
+
\omega_1'
| 1_\ell \rangle \langle 1_\ell |
\Big\}\;$,
with $\omega_1'$ a frequency close to $\omega_1$ satisfying $\omega_{p,s}-\omega_{p,r}=2\omega_1'$ \cite{Dimer2007}.
We then eliminate the excited states in the limit of large detuning $\Delta$ to finally obtain the multimode Dicke model
\eq{2Eq1}{\hat H=\sum_{i=1}^M \omega_i\hat a^\dag_{i}\hat{a}_{i}+\frac{\omega_z}{2}\sum_{l=1}^N \sigma^z_l+\sum_{i,l}\frac{g_{il}}{2}\sigma^x_l\left(\hat{a}^\dag_{i}+\hat{a}_{i}\right),}
with a correspondence of the effective spin operators in Eq.~(\ref{2Eq1}) 
to the internal atomic levels
\begin{align}
\sigma^z_\ell & =  | 1_\ell \rangle \langle 1_\ell | - | 0_\ell \rangle \langle 0_\ell |
\;,\;\;\;\;\;\;
\sigma^x_\ell  =  | 1_\ell \rangle \langle 0_\ell | + | 0_\ell \rangle \langle 1_\ell |\;.
\end{align}
The couplings and frequencies are tunable:
\begin{align}
\omega_i &= \nu_i - (\omega_{p,s}-\omega_1') + \frac{\tilde{g}_r^2(\mathbf{k}_i)}{\Delta_r}
\;,\;\;\;\;\;
\omega_z = 2(\omega_1-\omega_1'),
\nonumber\\
g_{i\ell} & = \frac{ g_r(\mathbf{k}_i,\mathbf{x}_\ell) \Omega_r}{2\Delta_r}\;,
\end{align}
where we assume Eq. (15) of Ref.~\cite{Dimer2007} to be satisfied: $\frac{g_r^2}{\Delta_r}= \frac{g_s^2}{\Delta_s}$ and 
$\frac{g_r\Omega_r}{\Delta_r}=\frac{g_s\Omega_s}{\Delta_s}$.
In particular the effective spin-photon coupling $g_{i\ell}$ can now be tuned sufficiently strong to reach superradiant regimes by changing the amplitude of the pump $\Omega_r$. The effective cavity frequencies receive an additional shift from a mode mixing term $a_{i}a_{j}$ with space averaged cavity couplings $\sim \tilde{g}^2/\Delta_r$ from which we only keep the mode-diagonal contribution (for running wave cavity mode functions $\sim e^{i\mathbf{k}_i \mathbf{x}_\ell}$ this is exact; we do not expect qualitative changes to our results from this approximation).

The multimode Dicke model with internal atomic levels obeys the same Ising-type $\mathbb{Z}_2$ symmetry, ${(\ann{a}{i},\sigma^x_l)\rightarrow (-\ann{a}{i},-\sigma_l^x)}$, familiar from the single-mode Dicke model \cite{Hepp1973, Wang1973, Carmichael1973, Cromer1974, Emary2003a, Emary2003b}. Therefore, there exists a critical coupling strength $J_c$, such that the ground state of the system spontaneously breaks the $\mathbb{Z}_2$ symmetry as soon as the average coupling strength
\eq{2Eq2}{
J\equiv \frac{1}{N}\sum_{l,m=1}^N\sum_{i=1}^M \frac{g_{il}g_{im}}{4}}
exceeds the critical value, ${J\geq J_c}$. The phase transition from the symmetric to the symmetry broken, superradiant (SR) phase has been well analyzed for the single-mode Dicke model and the essential findings, such as the universal behavior for zero and finite temperature transitions \cite{Emary2003a, Emary2003b} or in the presence of dissipation \cite{nagy11, oztop2012, Dalla2012}, remain valid also for the multimode case. The superradiant phase is determined by the presence of a photon condensate, i.e. the emergence of a coherent intra-cavity field \cite{baumann2010, Baumann2011}, which is described by a finite expectation value of a photon creation operator ${\langle \cre{a}{\mbox{\tiny C}}\rangle\neq 0}$. The superradiant condensate 
$
\cre{a}{\mbox{\tiny C}}=\sum_i \alpha_i^{\mbox{\tiny C}} \cre{a}{i},\ \ \mbox{with}\ \  \sum_i |\alpha_i^{\mbox{\tiny C}}|^2=1,$ is a superposition of many cavity modes $\cre{a}{i}$, and its explicit structure depends on the realization of the couplings $\{g_{il}\}$.

\subsection{Markovian dissipation}
In a cavity QED experiment of the type described in Fig.~\ref{fig:cavity}, the atoms and photons governed by the Hamiltonian \eqref{2Eq1} are additionally coupled to the electromagnetic field outside the cavity. This leads to the additional processes of spontaneously emitted photons into the environment and to cavity photon loss through imperfect mirrors, accurately captured by a Markovian master equation \cite{GardinerBook, CarmichaelBook}) of the form 
\eq{2Eq5}{\partial_t\rho=-i[\hat{H},\rho]+\mathcal{L}(\rho)\equiv\mathcal{L}\sub{tot}(\rho),}
where $\rho$ is the density matrix of the atom-photon system, $\hat H$ is the Hamiltonian \eqref{2Eq1} and $\mathcal{L}$ is a Liouville operator in Lindblad form
\eq{2Eq6}{\mathcal{L}(\rho)=\sum_{\alpha} \kappa_{\alpha}\left(2\ann{L}{\alpha}\rho \cre{L}{\alpha}-\{\cre{L}{\alpha}\ann{L}{\alpha},\rho\}\right).}
Here, ${\{\cdot\ ,\cdot\}}$ represents the anti-commutator and the $\ann{L}{\alpha}$ are Lindblad or quantum jump operators. The photon dissipation is described by the Liouvillian 
\eq{2Eq7}{\mathcal{L}\sub{ph}(\rho)=\sum_{i=1}^M \kappa_i \left(2\hat a_{i}\rho\hat{a}^\dag_{i}-\{\hat{a}^\dag_{i}\hat{a}_{i},\rho\}\right),}
where $\kappa_i$ is the loss rate of a cavity photon from mode $(i)$.
Eq.~(\ref{2Eq7}) describes a Markovian loss process that, while being a standard approximation in quantum optics, 
violates detailed balance between the system and the bath. Formally, it can be derived by starting with a cavity-bath setup 
in which both are at equilibrium with each other, and performing the transformation into the rotating frame 
outlined above Eq.~(\ref{2Eq1}) also on the system-bath couplings (see App. \ref{sec:markov}).

In this work, we consider ${\kappa_i<\omega_i,\omega_z}$ but of the same order of magnitude. In contrast, the atomic dissipative dynamics are considered to happen by far on the largest time scale, which can be achieved in typical cavity experiments \cite{baumann2010, Baumann2011}. In a recent open system realization of the single-mode Dicke model \cite{baumann2010, Baumann2011}, spontaneous individual atom-light scattering is suppressed by five orders of magnitude compared to the relevant system time-scales, such that atomic dephasing effectively plays no role \cite{baumann2010}. As a result, only global atomic loss is influencing the dynamics, which, however, can be compensated experimentally by steadily increasing the pump intensity or chirping the pump-cavity detuning \cite{baumann2010}. We therefore do not consider atomic spontaneous emission in this paper.

\subsection{Quenched / quasi-static disorder}\label{sec:QD}

The glassy physics addressed in this paper arises when the spatial variation of cavity mode couplings
\eq{2Eq3}{
K=\frac{1}{N}\sum_{l,m=1}^N \left(\sum_{i=1}^M \frac{g_{il}g_{im}}{4}\right)^2-J^2,}
is sufficiently large. The specific values of the couplings $g_{il}$ in Eq.~\eqref{2Eq1} are fluctuating as a function of the atom $(l)$ and photon $(i)$ numbers and depend on the cavity geometry and realization of the random trapping potential (Fig.~\ref{fig:cavity}). 
After integrating out the photonic degrees of freedom in Eq.~\eqref{2Eq1}, we obtain the effective atomic Hamiltonian 
\eq{2Extra1}{
H\sub{eff}=\frac{\omega_z}{2}\sum_{l=1}^N \sigma_l^z-\sum_{l,m=1}^N J_{lm} \sigma_l^x\sigma_m^x,}
where we introduced the effective atom-atom couplings
$J_{lm}=\sum_{i=1}^M \frac{g_{li}g_{im}}{4}$,
and at this point neglected the frequency dependence in the atom-atom coupling term in Eq.~\eqref{2Extra1}. This is appropriate for ${\omega_i\approx\omega_0}$ and $\omega_0$ large compared to other energy scales (in particular, $|(\omega_i - \omega_0)/\omega_0|\ll 1$).
In order to solve the effective Hamiltonian \eqref{2Extra1}, it is sufficient to know the distribution of the couplings $J_{lm}$, which itself is a sum over $M$ random variables. For a large number of modes (${M\rightarrow\infty}$), this distribution becomes Gaussian, according to the central limit theorem, with expectation value $J$ and variance $K$, as defined in Eqs.~\eqref{2Eq2},~\eqref{2Eq3}, respectively. 

The variables $J_{lm}$ can be seen as spatially fluctuating but temporally static variables, connecting all atoms with each other. This may be seen as a coupling to a bath with random variables $J_{lm}$, which vary on time scales $\tau_Q$ much larger than the typical time scales of the system $\tau_S$ only. The dynamics of the bath is then frozen on time scales of the system, and the bath is denoted as quasi-static or \emph{quenched} \cite{BinderYoung}.
This type of bath is in a regime opposite to a Markovian bath, where the dynamics of the bath happens on much faster time scales $\tau_M$ than for the system, ${\tau_M\ll\tau_S}$ \cite{GardinerBook, CarmichaelBook}. We have summarized basic properties of these baths in App.~\ref{sec:MvsQ}.

\section{Keldysh Path Integral Approach}\label{sec:pathintegral}
In this section, we introduce the Keldysh formalism \cite{Kamenev2009, KamenevBook,Dalla2012} and derive the set of self-consistency equations for the atoms and photons 
from which all our results can be extracted. We first formulate the open multi-mode Dicke model Eqs.~(\ref{2Eq1},\ref{2Eq7}) 
as an equivalent Keldysh action that includes the non-unitary time evolution induced by cavity decay.
In the Keldysh approach, one additionally benefits from the fact that the partition function \eq{2Eq8}{Z=\mbox{Tr}\left(\rho(t)\right)=1} 
is normalized to unity, independent of the specific realization of disorder, and we perform the disorder average directly on the partition function. We then integrate out the photons (carefully keeping track of their correlations, as explained below) and derive a set of saddle-point equations for frequency-dependent correlation functions which can be solved.

\subsection{Multi-mode Dicke action}\label{sec:ActionD}
To describe the photon dynamics, one starts from an action for the coupled system of cavity photons and a Markovian bath. Then the bath variables are integrated out in Born-Markov and rotating wave approximations.
The resulting Markovian dissipative action for the photonic degrees of freedom on the $(\pm)$-contour reads
\begin{eqnarray}
S\sub{ph}&=&\sum_j \int_{-\infty}^{\infty} dt\left(\crea{a}{j+}(i\partial_t-\omega_j)\ann{a}{j+}-(\crea{a}{j-}(i\partial_t-\omega_j)\ann{a}{j-}\right .\nonumber\\
&-&\left .i\kappa[2\ann{a}{j+}\crea{a}{j-}-(\crea{a}{j+}\ann{a}{j+}+\crea{a}{j-}\ann{a}{j-})]\right).\label{2Eq11}
\end{eqnarray}
Here, the creation and annihilation operators have been replaced by time-dependent complex fields. The structure of the master equation \eqref{2Eq5} is clearly reflected in the action on the $(\pm)$-contour in Eq.~\eqref{2Eq11}. The first line corresponds to the Hamiltonian part of the dynamics, with a relative minus sign between $(+)$ and $(-)$ contour stemming from the commutator. The second line displays the characteristic form of the dissipative part in Lindblad form.

For practical calculations, it is more convenient to switch from a $(\pm)$-representation of the path integral to the so-called Keldysh or RAK representation. In the latter, the fields on the $(\pm)$-contour are transformed to ``classical'' ${\ann{a}{j,c}=(\ann{a}{j+}+\ann{a}{j-})/\sqrt{2}}$ and ``quantum'' fields ${\ann{a}{j,q}=(\ann{a}{j+}-\ann{a}{j-})/\sqrt{2}}$, where the labeling of these fields indicates that $\ann{a}{j,c}$ can acquire a finite expectation value, while the expectation value of $\ann{a}{j,q}$ is always zero. After a Fourier transformation to frequency space, 
$\ann{a}{i}(\omega)=\int dt\ \ann{a}{i}(t)\ e^{-i\omega t}$,
the photonic action in Keldysh representation is obtained as
\eq{2Eq12}{
S\sub{ph}=\int_{j,\omega} (\crea{a}{j,c},\crea{a}{j,q}) \left(\begin{array}{cc} 0 & D_j^R(\omega)\\  
\phantom{l}D_j^A(\omega) & D_j^K(\omega)\end{array}\right) \left(\begin{array}{c} \ann{a}{j,c}\\ \ann{a}{j,q}\end{array}\right), }
where we used the abbreviation ${\int_{j,\omega}=\sum_{j=1}^M\intf}$. The integral kernel of Eq.~\eqref{2Eq12} is the inverse Green's function in Keldysh space with the inverse retarded/advanced Green's function
\eq{2Eq13}{
D^{R/A}_j(\omega)=[G^{R/A}_j]^{-1}(\omega)=\omega\pm i\kappa_j-\omega_j}
and the Keldysh component of the inverse Green's function
\eq{2Eq14}{
D^K_j(\omega)=2i\kappa_j.}
From now on, we will focus on the case where the variation in the photon parameters 
is much smaller than all other energy scales of this problem and consider only a single photon frequency $\omega_0$ and photon loss rate $\kappa$, i.e. ${|\kappa-\kappa_j|\ll \kappa}$ and ${|\omega_0-\omega_j|\ll\omega_0}$ for all photon modes $(j)$. As a result all photon Green's functions are identical with ${\kappa_j=\kappa}$ and ${\omega_j=\omega_0}$.
The Green's function in Keldysh space takes the form
\eq{2Eq15}{
\mathcal{G}(\omega)=\left(\begin{array}{cc} G^K(\omega) & G^R(\omega)\\ G^A(\omega) & 0\end{array}\right)=\left(\begin{array}{cc} 0 & D^R(\omega)\\  
\phantom{l}D^A(\omega) & D^K(\omega)\end{array}\right)^{-1},}
where we already identified retarded/advanced Green's function ${G^R(\omega)}$ in Eq.~\eqref{2Eq13}. The Keldysh component of the Green's function is obtained by performing the inversion \eqref{2Eq15} as
\eq{2Eq16}{
G^K(\omega)=-G^R(\omega)D^K(\omega)G^A(\omega).}
The retarded Green's function encodes the response of the system to external perturbations and its anti-hermitian part is proportional to the spectral density 
\eq{2Eq17}{
\mathcal{A}(\omega)=i\left(G^R(\omega)-G^A(\omega)\right),}
since ${G^A(\omega)=\left[G^R(\omega)\right]^{\dagger}}$.
The retarded Green's function ${G^R(\omega)}$ and the Keldysh Green's function ${G^K(\omega)}$ constitute the basic players in a non-equilibrium path integral description, determining the system's response and correlations.
For a more detailed discussion of a Keldysh path integral description of cavity photons, we refer the reader to \cite{Dalla2012}.

The atomic sector of the Dicke Hamiltonian \eqref{2Eq1} can be mapped to an action in terms of real fields $\phi_l$, as long as the physically relevant dynamics happens on frequencies below $\omega_z$ \cite{SachdevBook}. The $\phi_\ell$ obey the non-linear constraint
\eq{2Eq21}{
\delta(\phi_l^2(t)-1)=\int \mathcal{D}\lambda_l(t) e^{i\lambda_l(t)(\phi_l^2(t)-1)}\;,}
where ${\lambda_l(t)}$ are Lagrange multipliers, 
in order to represent Ising spin variables (see Ref.~\onlinecite{Dalla2012} for further explanation). As a result, 
we can apply the following mapping to Eq.~(\ref{2Eq1}) 
\begin{eqnarray}
\sigma^x_l(t)&\longrightarrow &\phi_l(t),\label{2Eq19}\\
\sigma^z_l(t)&\longrightarrow &\frac{2}{\omega_z^2} \left(\partial_t\phi_l(t)\right)^2-1,\label{2Eq20}
\end{eqnarray}
On the $(\pm)$-contour, we then obtain
\eq{2Eq22}{
S\sub{at}=\frac{1}{\omega_z}\int_{l,t} \left(\partial_t\phi_{l+}\right)^2-\left(\partial_t\phi_{l-}\right)^2,}
subject to the non-linear constraint
\eq{2Eq23}{
S\sub{const}=\frac{1}{\omega_z}\int_{l,t} \lambda_{l+}\left(\phi_{l+}^2-1\right)-\lambda_{l-}\left(\phi_{l-}^2-1\right).}
The atom-photon coupling reads
\eq{2Eq24}{
S\sub{coup}=\int_{t,i,l} \frac{g_{il}}{2} \left(\phi_{l+}\left(\crea{a}{i+}+\ann{a}{i+}\right)-\phi_{l-}\left(\crea{a}{i-}+\ann{a}{i-}\right)\right).}
Transforming to the RAK basis and frequency space, the atomic propagator becomes
\begin{eqnarray}
S\sub{at}&=&\frac{1}{\omega_z}\int_{\omega,l}\left(\phi_{c,l},\phi_{q,l}\right)D\sub{at}(\omega)\left(\begin{array}{c}\phi_{c,l}\\ \phi_{q,l}\end{array}\right)+\frac{1}{\omega_z}\int_{\omega,l}\lambda_{q,l}, \label{2Eq25}\end{eqnarray}
with
\eq{2Eq26}{
D\sub{at}(\omega)=\left(\begin{array}{cc}\lambda_{q,l} & \lambda_{c,l}-\left(\omega+i\eta\right)^2\\\lambda_{c,l}-\left(\omega-i\eta\right)^2 & \lambda_{q,l}\end{array}\right).}
Here, ${\eta\rightarrow0^{+}}$ plays the role of a regulator that ensures causality for the retarded/advanced Green's functions. 
For the atom-photon coupling in the RAK basis, we get
\begin{eqnarray}
S\sub{coup}&=&\int_{\omega,l,j}\frac{g_{il}}{2}\left(\left(\phi_{c,l},\phi_{q,l}\right)\sigma^x\left(\begin{array}{c}\ann{a}{c,l}\\ \ann{a}{q,l}\end{array}\right)\right . 
\left .+(\crea{a}{c,l},\crea{a}{q,l})\sigma^x\left(\begin{array}{c}\phi_{c,l}\\ \phi_{q,l}\end{array}\right)  \right).
\nonumber\\
\label{2Eq27}
\end{eqnarray}
For the atomic fields, it is useful in the following to introduce the Keldysh vector
$\Phi_l(\omega)=\left(\begin{array}{c}\phi_{c,l}(\omega)\\ \phi_{q,l}(\omega)\end{array}\right)$,
which will simplify the notation in the following.

The Keldysh action for the open multimode Dicke model is then obtained as the sum of Eqs.~(\ref{2Eq12},\ref{2Eq25}, \ref{2Eq27})
\eq{2Eq28}{
S\left[\{ \cre{a}{}, \ann{a}{}, \phi, \lambda\}\right]=S\sub{ph}+S\sub{at}+S\sub{coup}.}

\subsection{Calculation procedure}\label{sec:calc}

\begin{table*}
\begin{tabular}{ |l |l ||l| }
  \hline
  \multicolumn{2}{|c||}{\rule{0mm}{2ex}Atoms}&\multicolumn{1}{c|}{\rule{0mm}{2ex} Photons} \\
  \hline\hline

  \rule{0mm}{4ex}

 & $~ Q_{cc}(t,t')=Q^K(t,t')=-i\ \overline{\left\langle \{\sigma^x_l(t),\sigma^x_l(t')\}\right\rangle}~$ &$ ~ G_{cc}(t,t')=G^K(t,t')=-i\ \overline{\left\langle \{\ann{a}{m}(t),\cre{a}{m}(t')\}\right\rangle}~$\\
   $~ \psi_c(t)=\sqrt{2}\ \overline{\langle\sigma^x_l(t)\rangle}~$  & $~ Q_{cq}(t,t')=Q^R(t,t')=-i\ \Theta(t-t')\ \overline{\left\langle [\sigma^x_l(t),\sigma^x_l(t')]\right\rangle} ~$& $~ G_{cq}(t,t')=G^R(t,t')=-i\ \Theta(t-t')\ \overline{\left\langle [\ann{a}{m}(t),\cre{a}{m}(t')]\right\rangle} ~$ \\
  $~\psi_q(t)=0$ & $~ Q_{qc}(t,t')=Q^A(t,t')=\left(Q^R(t,t')\right)^{\dagger} ~$& $~ G_{qc}(t,t')=G^A(t,t')=\left(G^R(t,t')\right)^{\dagger} ~$\\
   & $~ Q_{qq}(t,t')=0~$& $~ G_{qq}(t,t')=0~$ \\[1ex]
  \hline
\end{tabular}
\caption{Translation table for the atomic order parameter and Green's functions, from now on labeled with $Q$, and the intra-cavity photon Green's function, labeled with $G$.}
\end{table*}
We now explain how we solve the Keldysh field theory described by Eq.~(\ref{2Eq28}). The calculation proceeds in three steps: 

\emph{1. Integration of the photon modes:} This step can be performed exactly via Gaussian integration, since the action \eqref{2Eq28} is quadratic in the photon fields. Note that this does not mean that we discard the photon dynamics from our analysis. To also keep track of the photonic observables, we modify the bare inverse photon propagator, Eq.~\eqref{2Eq12}, by adding (two-particle) source fields $\mu$ according to
\eq{2Eq29}{
D\sub{ph}(\omega)\rightarrow D\sub{ph}(\omega)+\mu(\omega), \ \ \mbox{with}\ \ \mu=\left(\begin{array}{cc}\mu^{cc}&\mu^{cq}\\ \mu^{qc}&\mu^{qq}\end{array}\right).}
The photon Green's functions are then obtained via functional variation of the partition function with respect to the source fields
\eq{2Eq30}{
G^{R/A/K}(\omega)=\left .\frac{\delta}{\delta\mu^{qc/cq/cc}(\omega)}Z\right|_{\mu=0}.}The resulting action is a sum of the bare atomic part \eqref{2Eq25} and an effective atom-atom interaction
\eq{2Eq31}{
S\sub{at-at}=-\int_{\omega}\sum_{l,m}J_{lm}\Phi^{T}_l(-\omega)\Lambda(\omega)\ann{\Phi}{m}(\omega),}
with atom-atom coupling constants $J_{lm}$ defined in \eqref{2Extra1} and the frequency dependent coupling 
\eq{2Eq32}{
\Lambda(\omega)=\frac{1}{2}\sigma^x\left(G^{\phantom{T}}_0(\omega)+G^{T}_0(-\omega)\right)\sigma^x,}
which is the bare photon Green's function $G_0$ after symmetrization respecting the real nature of the Ising fields $\Phi_l$. We note that the information of the photonic coupling to the Markovian bath is encoded in ${\Lambda(\omega)}$. 

\emph{2. Disorder average:} The coupling parameters $J_{lm}$ are considered to be Gaussian distributed and the corresponding distribution function is determined by the expectation value and covariance of the parameters $J_{lm}$
\begin{eqnarray}
\overline{J_{lm}}&=&\frac{J}{N},\\ \overline{\delta J_{lm}\delta J_{l'm'}}&=&\frac{K}{N}\left(\delta_{ll'}\delta_{mm'}+\delta_{lm'}\delta_{ml'}\right)\equiv \hat{K}_{lml'm'}\ ,\phantom{MM}\end{eqnarray}
where the line denotes the disorder average and ${\delta J_{lm}=J_{lm}-\overline{J_{lm}}}$ represents the variation from the mean value. The disorder averaged partition function can be expressed as
\eq{2Eq33}{
\overline{Z}=\int \mathcal{D}\left(\{\Phi,\lambda,J\}\right) e^{i(S\sub{at}+S\sub{at-at}+S\sub{dis})},}
with the disorder ``action'' 
\eq{2Eq34}{
S\sub{dis}= \frac{i}{2}\sum_{l,m,l',m'}\left(J_{lm}-\overline{J_{lm}}\right)\hat{K}^{-1}_{lml'm'}\left(J_{l'm'}-\overline{J_{l'm'}}\right),}
describing a temporally frozen bath with variables $J_{lm}$.
Performing the disorder average, i.e. integrating out the variables $J_{lm}$ in the action \eqref{2Eq33} replaces the parameters ${J_{lm}\rightarrow J/N}$ in \eqref{2Eq31}
by their mean value. Furthermore, the variance $K$ introduces a quartic interaction term for the atomic Ising variables which is long-range in space 
\eq{2Eq35}{
S\sub{at-4}=\frac{iK}{N}\int_{\omega,\omega'}\sum_{l,m}\left(\Phi_l\Lambda\Phi_m\right)(\omega)\left(\Phi_l\Lambda\Phi_m\right)(\omega'),}
with the shortcut 
$(\Phi_l\Lambda\Phi_m)(\omega)\equiv \Phi_l^{T}(\omega)\Lambda(\omega)\Phi_m(\omega)$.

\emph{3. Collective variables: Atomic order parameter and Green's function:} To decouple the spatially non-local terms in \eqref{2Eq31} and \eqref{2Eq35}, we introduce the Hubbard-Stratonovich fields
$\psi_{\alpha}$ and $Q_{\alpha\alpha'}$ with ${\alpha,\alpha'=c,q}$, which represent the atomic order parameter
\eq{2Eq36}{
\psi_{\alpha}(\omega)=\frac{1}{N}\sum_l\langle\overline{ \phi_{\alpha,l}(\omega)\rangle}}
and average atomic Green's function
\eq{2Eq37}{
Q_{\alpha\alpha'}(\omega,\omega')=\frac{1}{N}\sum_l\overline{ \langle \phi_{\alpha,l}(\omega)\phi_{\alpha',l}(\omega')\rangle}.}
Now, the action is quadratic in the original atomic fields $\phi_\ell$, and so these can be integrated out. The resulting action has a global prefactor $N$ and we will perform a saddle-point approximation which becomes exact in the thermodynamic limit and upon neglecting fluctuations of the Lagrange multiplier. We replace the fluctuating Lagrange multipliers ${\lambda_l(t)}$ by their saddle-point value ${\lambda_l(t)=\lambda}$. In the steady state, the atomic observables become time-translational invariant which restricts the frequency dependence of the fields to
\begin{eqnarray}
\psi_{\alpha}(\omega)&=&2\pi \delta(\omega)\psi_{\alpha},\label{eq:ferro}\\
Q_{\alpha\alpha'}(\omega,\omega')&=&2\pi \delta(\omega+\omega') Q_{\alpha\alpha'}(\omega).
\end{eqnarray}\\

\subsection{Saddle-point action and self-consistency equations}
The saddle-point action is given by the expression
\begin{widetext}
\eq{2Eq38}{
\mathcal{S}/N=-\frac{2\lambda_q}{\omega_z}+\int_{\omega}\Psi^{T}(-\omega)\left[J\Lambda(\omega)-J^2\Lambda(\omega)\tilde{G}(\omega)\Lambda(\omega)\right]\Psi(\omega)-\frac{i}{2}\mbox{Tr}\left[\ln \tilde{G}(\omega)\right]+iK\mbox{Tr}\left[\Lambda Q\Lambda Q\right] (\omega),}
\end{widetext}
with the ``Green's function''
\eq{2Eq39}{
\tilde{G}(\omega)=\left(D^{\phantom{\int}}\sub{at}(\omega)-2K\Lambda(\omega) Q(\omega) \Lambda(\omega)\right)}
and the field ${\Psi^T=(\psi_{c},\psi_q)}$. The matrices $\Lambda, \tilde{G}$ and $Q$ in Eq.~\eqref{2Eq38} possess Keldysh structure, i.e. they are frequency dependent triangular matrices with retarded, advanced and Keldysh components. The matrix $\Lambda$ contains the photon frequencies $\omega_0$, the decay rate $\kappa$, and also depends on the photon Lagrange multiplier $\mu$, so that all photon correlations can be extracted from Eq.~\eqref{2Eq38}.
\subsubsection{Atomic sector}
 In order to find a closed expression for the macroscopic fields $\{\Phi, Q\}$ and to determine the saddle-point value for the Lagrange multiplier $\lambda$, we have to evaluate the saddle-point equations
\eq{2Eq40}{
\frac{\delta S}{\delta X}\overset{!}{=}0, \ \ \ \mbox{with} \ X=Q_{\alpha\alpha'},\psi_{\alpha}, \lambda_{\alpha}, \ \ \alpha=c,q.}
In stationary state, ${\lambda_q=\psi_q=Q_{qq}=0}$ by causality and we set ${\lambda_{c}=\lambda}$ and ${\psi_{c}=\psi}$ for convenience.

The saddle-point equation for $\lambda_q$ expresses the constraint
\eq{2Eq41}{
2=\int_{\omega}iQ^K(\omega)=iQ^K(t=0)=2\frac{1}{N}\sum_{l=1}^N\langle\overline{ (\sigma_l^x)^2\rangle},}
which has been reduced to a soft constraint, present on average with respect to $(l)$, compared to the original hard constraint, ${(\sigma_l^x)^2=1}$ for each $(l)$ individually. 

In the superradiant phase and in the glass phase, the spin attain locally ``frozen'' configurations. The correlation time of the system becomes infinite, expressed via a non-zero value of the Edwards-Anderson order parameter
\eq{2Eq4}{q\sub{EA}:=\lim_{\tau\rightarrow\infty}\frac{1}{N}\sum_{l=1}^N\overline{\langle\sigma^x_l(\tau)\sigma^x_l(0)\rangle}.}
As a consequence, the correlation function $Q^K(\omega)$ is the sum of a regular part, 
describing the short time correlations and a $\delta$-function at $\omega=0$, caused by the infinite correlation time. We decompose the correlation function according to
\eq{2Eq42}{
Q^K(\omega)=4i\pi q\sub{EA} \delta(\omega)+Q^K\sub{reg}(\omega),}
with the Edwards-Anderson order parameter $q\sub{EA}$, being defined in Eq.~\eqref{2Eq4} and a regular contribution $Q^K\sub{reg}$. In the literature \cite{Cugliandolo1999, Kennett2001}, this decomposition is referred to as modified fluctuation dissipation relation (FDR) as also discussed in Appendix~\ref{sec:MvsQ}.
The saddle-point equations for atomic response function and the regular part of the Keldysh function are
\eq{2Eq47}{
Q^R(\omega)=\left(\frac{2(\lambda-\omega^2)}{\omega_z}-4K\left(\Lambda^R(\omega)\right)^2Q^R(\omega)\right)^{-1}}
and 
\eq{2Eq48}{
Q^K\sub{reg}(\omega)=\frac{4K\left|Q^R\right|^2\Lambda^K\left(Q^A\Lambda^A+Q^R\Lambda^R\right)}{1-4K\left|Q^R\Lambda^R\right|^2}.}
Eqs.~\eqref{2Eq41}, \eqref{2Eq47}, \eqref{2Eq48} form a closed set of non-linear equations, describing the physics of the atomic subsystem in the thermodynamic limit, which will be discussed in Sec.~\ref{sec:Results}. 

\subsubsection{Photonic sector}
The photon response $G^R$ and correlation function $G^K$ are determined via functional derivatives of the partition function $\mathcal{Z}$ with respect to the source fields $\mu$, as described in \eqref{2Eq29} and \eqref{2Eq30}. The saddle-point for the partition function is
\eq{2Eq49}{
\mathcal{Z}=e^{i\mathcal{S}}\times Z^{(0)}\sub{ph},}
with the action $\mathcal{S}$ from Eq.~\eqref{2Eq38} and the bare photon partition function $Z^{(0)}\sub{ph}$.

 In the Dicke model, the photon occupation $n_i$ is not a conserved quantity, such that anomalous expectation values
${\left\langle a^2\right\rangle\neq0}$ will become important. This has to be taken into account by introducing a Nambu representation, where the photon Green's functions become ${2\times2}$ matrices, see Appendix~\ref{app:Nambu}. 
Generalizing the source fields $\mu$ to include normal and anomalous contributions, and evaluating the functional derivatives with respect to these fields, results in the inverse photon response function
\begin{eqnarray}
& &D_{2\times2}^R(\omega)=\\
& &\left(\begin{array}{cc} \omega+i\kappa-\omega_0+\Sigma^R(\omega)&\Sigma^R(\omega)\\ \left(\Sigma^R(-\omega)\right)^{*}&-\omega-i\kappa-\omega_0+\left(\Sigma^R(-\omega)\right)^{*}\end{array}\right).\label{eq:DR_photon}
\end{eqnarray}
Here, the subscript $2\times2$ indicates Nambu representation and
\eq{2Eq52}{\Sigma^R(\omega)=\left(\Sigma^R(-\omega)\right)^{*}=\frac{1}{2\Lambda^R(\omega)}\left(\frac{\omega_zD^R\sub{at}(\omega)}{2\left(\omega^2-\lambda\right)}-1\right)}
is the self-energy, resulting from the atom-photon interaction. The Keldysh component of the inverse Green's function is
\eq{2Eq53}{
D_{2\times2}^K(\omega)=\left(\begin{array}{cc} 2i\kappa+\Sigma^K(\omega) &\Sigma^K(\omega)\\-\left(\Sigma^K(\omega)\right)^{*} &2i\kappa-\left(\Sigma^K(\omega)\right)^{*}\end{array}\right)}
with the self-energy
\eq{2Eq55}{\Sigma^K(\omega)=-\left(\Sigma^K(\omega)\right)^{*}=\frac{Q^K(\omega)}{4\mbox{Re}\left(Q^R(\omega)\Lambda^R(\omega)\right)}.}

In the Dicke model, the natural choice of representation for the photon degrees of freedom is the {$x$-$p$} basis, i.e. in terms of the real fields 
$x=\frac{1}{\sqrt{2}}(\crea{a}{}+a), \ \ p=\frac{1}{\sqrt{2}i}(\crea{a}{}-a)$,
since the atom-photon interaction couples the photonic $x$-operator to the atoms. In this basis, the self-energy gives only a contribution to the $x$-$x$ component of the inverse Green's function, and the inverse response function reads
\eq{2Eq57}{
D_{xp}^R(\omega)=\left(\begin{array}{cc} 2\Sigma^R(\omega)-\omega_0 &\kappa-i\omega\\ -\kappa+i\omega &-\omega_0\end{array}\right).}
In the limit of vanishing disorder ${K\rightarrow 0}$, the self-energy approaches the value
$\Sigma^R(\omega)=\frac{J\omega_z}{2\left(\omega^2-\lambda\right)}$,
reproducing the result for the single mode Dicke model \cite{Dalla2012, Emary2003a, Emary2003b}.

\section{Results}\label{sec:Results}

We now present our predictions from solving the atomic saddle-point equations Eqs.~\eqref{2Eq41}, \eqref{2Eq47}, \eqref{2Eq48} and then extracting the photonics correlations using Eqs.~\eqref{eq:DR_photon}-\eqref{2Eq57}, in the same order as in the Key Results Section \ref{sec:keyresults}. In the subsection {\it Cavity Glass Microscope}, we present signatures for standard experimental observables of cavity QED by adapting the input-output formalism to the Keldysh path integral. 

\subsection{Non-equilibrium steady state phase diagram}

The phases in the multimode Dicke model shown in Fig.~\ref{fig:PhaseDiag} can be distinguished by the two order parameters, namely the Edwards-Anderson order parameters $q\sub{EA}$  in Eq.~\eqref{2Eq4}, indicating an infinite correlation time $\tau$ and the ferromagnetic order parameter $\psi$ defined (Eq.~(\ref{eq:ferro})), indicating a global magnetization:
\begin{center} \begin{tabular}{|c||c|c|}
\hline Normal & $\ q\sub{EA}=0\ $ &  $\ \psi=0\ $  \\ 
\hline SR & $\ q\sub{EA}\neq0\ $ &  $\ \psi\neq0\ $ \\ 
\hline QG & $\ q\sub{EA}\neq0\ $ &  $\ \psi=0\ $ \\ 
\hline \end{tabular} \end{center}
In the normal phase, the Edwards-Anderson parameter $q\sub{EA}$ and the ferromagnetic order parameter $\psi$ are both zero and Eq.~\eqref{2Eq41} implicitly determines the numerical value of the Lagrange parameter $\lambda\sub{N}$. In contrast, in the superradiant phase ${\psi\neq 0}$, and the Lagrange parameter can be determined analytically to be
\eq{2Eq43}{
\lambda\sub{SR}=\frac{\omega_z\omega_0}{\omega_0^2+\kappa^2}\left(J+\frac{K}{J}\right).}
In the quantum glass phase the Lagrange parameter is pinned to
\eq{2Eq44}{
\lambda\sub{QG}=\frac{\omega_z\omega_0}{\omega_0^2+\kappa^2}\sqrt{K}.}
The normal phase is characterized by a vanishing Edwards-Anderson order parameter, and the corresponding Lagrange multiplier $\lambda\sub{N}$ is determined via the integral
\eq{2Eq45}{
0=2-\left .i\int_{\omega}Q\sub{reg}^K(\omega)\right|_{\lambda=\lambda\sub{N}}.}
The normal-SR phase border is located at the line for which $\lambda\sub{N}=\lambda\sub{SR}$, while the normal-QG transition happens at $\lambda\sub{N}=\lambda\sub{QG}$. In the same way, the transition between superradiant phase and quantum glass phase happens when $\psi$ vanishes for finite ${q\sub{EA}\neq0}$. This is the case for
\eq{2Eq46}{
\lambda\sub{SR}=\lambda\sub{QG}\Leftrightarrow K=J^2.}
The phase diagram for the open system for different values of the photon dissipation $\kappa$ is shown in Fig.~\ref{fig:PhaseDiag}. As can be seen from this figure, the qualitative features of the zero temperature phase diagram \cite{Strack2011} are preserved in the presence of dissipation. However, with increasing $\kappa$, the phase boundaries between normal and SR, QG phase are shifted to larger values of $J$, $K$ respectively, while the QG-SR transition is still located at the values for which ${J^2=K}$ as for the zero temperature equilibrium case. Finite dissipation neither favors the QG nor the SR phase and as a result, the competition between disorder and order is not influenced by the dissipative dynamics. The line at ${K=0}$, i.e. zero disorder, describes the normal-SR transition for the single mode Dicke model, which is known to be located at ${J_c=\frac{\omega_0^2+\kappa^2}{4\omega_0}\omega_z}$ \cite{Dalla2012, Emary2003a, Emary2003b}. This result is exactly reproduced within our approach.

\subsection{Dissipative spectral properties and universality class}\label{sec:AtSpec}

\begin{figure}[t!]
\includegraphics[width=1\linewidth]{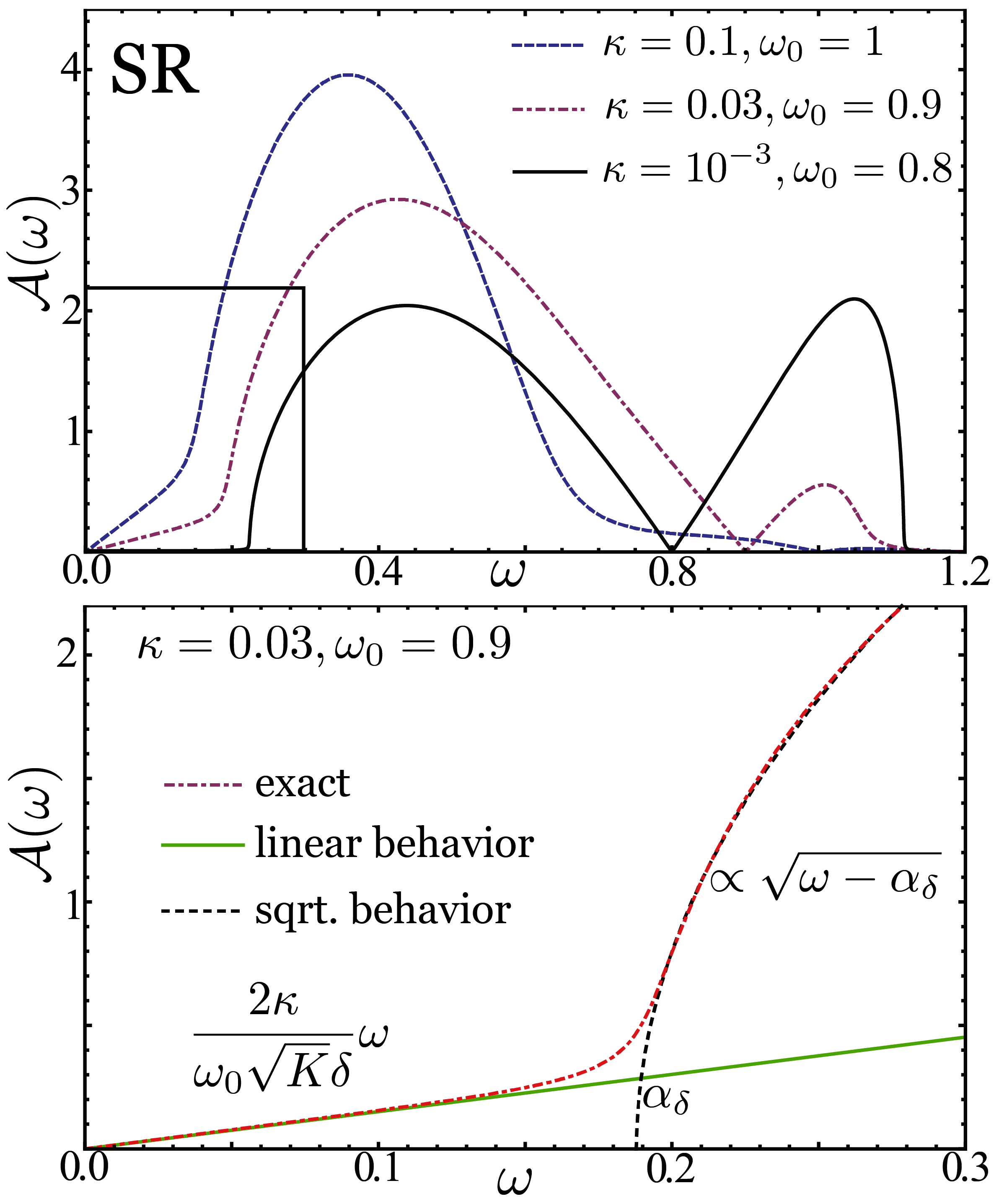}
    \caption
    {\label{fig:FMSpec2}
    (Color online) Regular part of the spectral density ${\mathcal{A}(\omega)}$ in the superradiant phase for parameters ${K=0.05}$ and ${J=0.4}$ and varying $\kappa$ and $\omega_0$. For small frequencies ${\omega<\alpha_{\delta}}$ the spectral density is linear in $\omega$ and $\kappa$ and behaves as a square root for intermediate frequencies ${\omega>\alpha_{\delta}}$. For the non-dissipative case (${\kappa\rightarrow0}$), the spectral weight develops a gap at low frequencies, which is indicated for ${\kappa=10^{-3}}$ (solid line). The lower panel depicts the low frequency behavior of $\mathcal{A}$ (red, dash-dotted line) for values ${\kappa=0.03}$ and ${\omega_0=0.9}$. The green (full) and the black (dashed) line  indicate the linear, square root behavior, respectively. Approaching the glass transition, $\alpha_{\delta}$ scales to zero ${\propto \delta^{\frac{3}{2}}}$.}
\end{figure}

The atomic excitation spectrum and the influence of the system-bath coupling on the atomic dynamics are encoded in the retarded atomic Green's function, which is identical to the atomic linear susceptibility, ${Q^R(\omega)=\chi^{(1)}(\omega)}$. It describes the response of the atomic system to a weak perturbation as, for instance, the coupling to a weak coherent light field (see appendix \ref{sec:Polarization}), and its imaginary part determines the atomic spectral response
\eq{4Extra1}{\mathcal{A}(\omega)=-2\mbox{Im}\left(Q^R(\omega)\right),} 
which can be measured directly via radio-frequency spectroscopy \cite{Stewart2008, Haussmann2009}. 
\begin{table}
\begin{tabular}{|C{4,5cm} |C{1.5cm} | C{1.9cm}| }
  \hline
  \rule{0mm}{3ex}
Analytic expressions & SR to QG&Normal to QG\\
  \hline\hline
\rule{0mm}{4ex} $Q^R(\omega)=Z_{\delta}\left(\left(\omega+i\gamma_{\delta}\right)^2-\alpha_{\delta}^2\right)^{-1}$& &\\
$\alpha_{\delta}=\frac{\sqrt{2}\left(\omega_0^2+\kappa^2\right)}{8\sqrt{K}^{3}\kappa}\ \times$ &$ \delta^{\frac{3}{2}}$&$\left|\frac{\delta}{\log (\delta)}\right|^{\frac{3}{2}}$ \\
$\gamma_{\delta}=\frac{\omega_0^2+\kappa^2}{16K^2\kappa}\times$& $ \delta^{2}$&$ \left|\frac{\delta}{\log (\delta)}\right|^{2}$ \\
$Z_{\delta}=\frac{\omega_0\left(\omega_0^2+\kappa^2\right)}{8\sqrt{K}^{5}\kappa^2}\times$ &$ \delta^3$&$\left|\frac{\delta}{\log (\delta)}\right|^3$ \\[2ex]
  \hline
\end{tabular}
\caption{Atomic spectral response and scaling behavior approaching the glass transition in two different ways. The normal to glass transition shows logarithmic corrections compared to the SR to QG transition. The logarithmic scaling correction is a typical feature of the glass transition and has also been found for $T=0$ and finite temperature glass transitions in equilibrium \cite{Ye1993,Miller1993}. We see that the life-time of the excitations $\gamma_{\delta}$ scales differently from the excitation energy $\alpha_{\delta}$, which indicates a strong competition of the reversible quantum dynamics and the classical relaxational dynamics. Although the inverse life-time scales faster to zero than the excitation energy, there is no point before the transition, where one of these quantities becomes exactly zero as it was the case for the superradiance transition. The described behavior at the glass transition means that there is no purely relaxational low energy theory which is able to describe the dynamics close to the transition. It does not fall into the Halperin-Hohenberg classification of dissipative dynamical systems, but belongs to the universality class of dissipative spin glasses \cite{Read1995,Cugliandolo1999,Cugliandolo2002,Cugliandolo2004}.}
\label{table:scaling}
\end{table}
The spectral response $\mathcal{A}$ for the normal and SR phase (the regular part for the latter) is shown in Fig.~\ref{fig:FMSpec2}.
In order to describe the low frequency behavior of the atomic spectrum, we decompose it into a regular and singular part, where the singular part captures the critical mode of the SR phase in terms of a $\delta$-function at zero frequency, which is absent in the normal phase. The regular part of the spectrum has the same structure for the normal and superradiant phase, and a derivative expansion of the inverse Green's function yields the low frequency response function
\eq{4Eq1}{
Q^R(\omega)=Z_{\delta}\left(\left(\omega+i\gamma_{\delta}\right)^2-\alpha_{\delta}^2\right)^{-1}}
with the analytic expressions for the coefficients given in Table \ref{table:scaling}.
This is the Green's function of a damped harmonic oscillator with characteristic frequency ${\omega=\alpha_{\delta}}$ and damping $\gamma_{\delta}$, which is described by classical relaxational dynamics and correctly determines the atomic spectral response for frequencies ${\omega<||\alpha_{\delta}-i\gamma_{\delta}||}$ smaller than the pole. The index $\delta$ in Eq.~\eqref{4Eq1} indicates that the parameters scale with the distance to the glass transition
\eq{4Eq2}{
\delta=K_c-K,}
which happens at ${K=K_c}$ (${\delta>0}$ in SR and normal phases). Transforming the response function to the time-domain,
\eq{4Eq3}{
Q^R(t)=Z_{\delta} e^{-\gamma_{\delta}t} \cos\left(\alpha_{\delta}t\right),} which describes an excitation of the system with inverse life-time ${\gamma_\delta=1/\tau_\delta}$, energy $\alpha_{\delta}$ and quasi-particle residuum $Z_{\delta}$. 
For frequencies ${\omega<\alpha_{\delta}}$, the spectral response is determined by the imaginary part of Eq. \eqref{4Eq1}, yielding
\eq{4Eq5}{\mathcal{A}(\omega)\approx \frac{Z_{\delta}\gamma_{\delta}\omega}{\alpha_{\delta}^4}=\frac{2\kappa }{\omega_0 \sqrt{K} \delta}\ \omega.}
This linear behavior is completely determined by parameters of the quenched and the Markovian bath and vanishes for ${\kappa\rightarrow0}$, resulting in a gap in the spectral weight for the zero temperature equilibrium case, as discussed in \cite{Strack2011}. For ${\omega>\alpha_{\delta}}$, i.e. for $\omega$ larger than the gap, the atomic response function develops a non-analytic square root behavior 
\eq{4Eq6}{
\mathcal{A}(\omega)\propto\sqrt{\omega-\alpha_{\delta}},}
resulting from the quadratic form in Eq.~\eqref{2Eq47}. The scaling of the excitation gap $\alpha_{\delta}$ and the ratio $\frac{Z_{\delta}\gamma_{\delta}}{\alpha_{\delta}^4}$ can be obtained directly from the atomic spectral response, as illustrated in Fig.~\ref{fig:FMSpec2}, lower panel. At the glass transition, the gap vanishes, such that the square root behavior starts from ${\omega=0}$.

\begin{figure}[t!]
\includegraphics[width=1\linewidth]{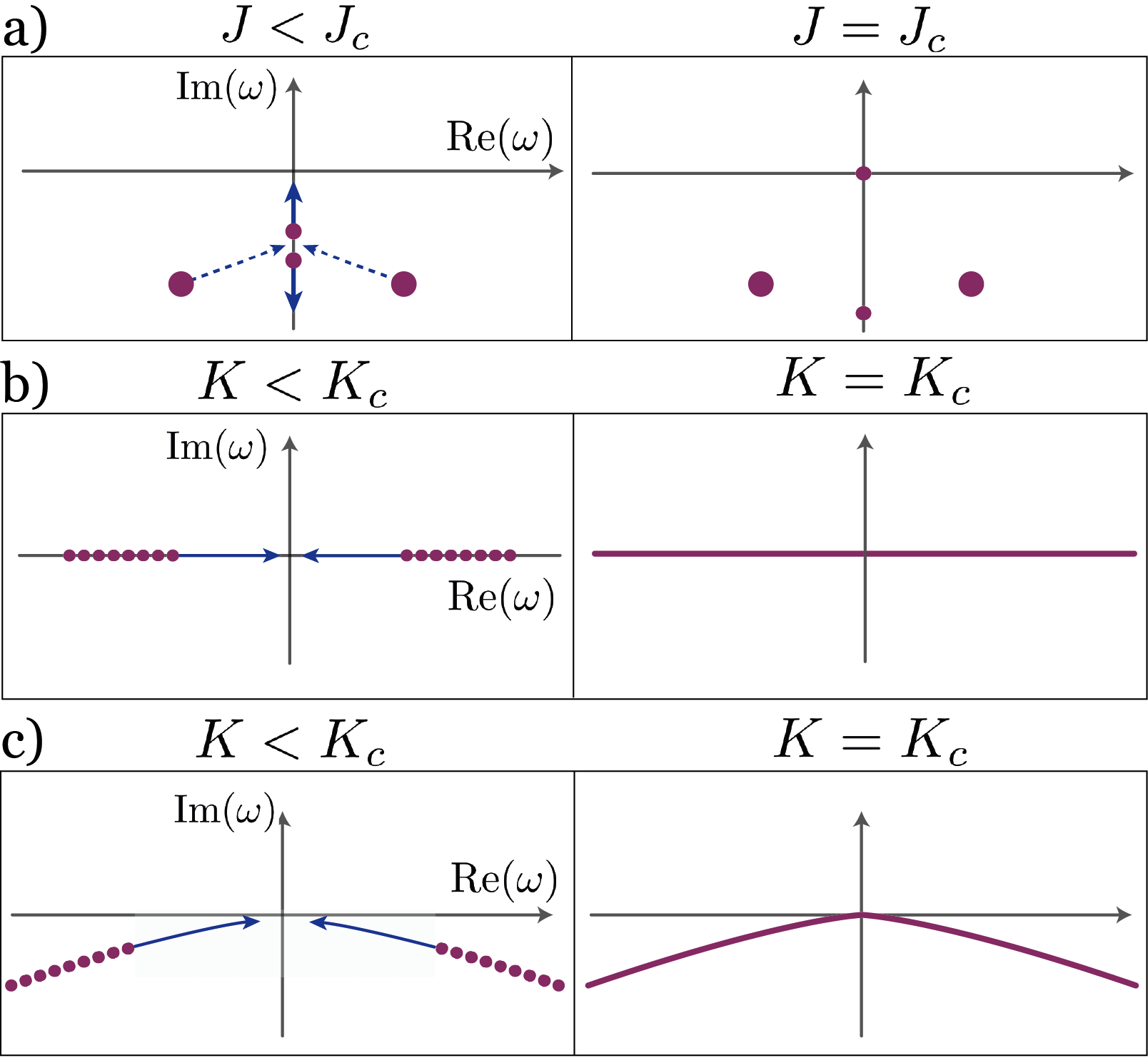}
    \caption
    {\label{fig:PoleStruct}
    (Color online) Schematic illustration of the pole structure and critical dynamics in the present model: a) the normal to superradiance transition in the dissipative Dicke model, b) the normal to glass transition in the zero temperature equilibrium model, c) the normal to glass transition in the dissipative model.\\
    a) When approaching the superradiance transition, two of the polaritonic modes advance to the imaginary axis and become purely imaginary before the transition point. This leads to the effective classical relaxational dynamics close to the transition. At the transition point, the $\mathbb{Z}_2$ symmetry is broken by only a single mode approaching zero and becoming critical for $J\rightarrow J_c$.\\
    b) For moderate disorder $K$, the poles are located on the real axis away from zero. For increasing $K$, the poles approach zero, with the closest pole scaling proportional to $|K-K_c|^{\frac{1}{2}}$.
 At $K=K_c$ the modes form a continuum which reaches zero and becomes quantum critical. No dissipative dynamics is involved.\\
    c) For moderate disorder $K$, the set of modes is located in the complex plane, away from zero. For increasing variance $K$, the modes get shifted closer to the origin, however, due to the scaling of real ($\propto |K-K_c|^{\frac{3}{2}}$) and imaginary part ($\propto|K-K_c|^{2}$), they neither become purely real nor purely imaginary. At $K=K_c$ a continuum of modes reaches zero.}
\end{figure}

In Table~\ref{table:scaling}, we compare the scaling behavior of the atomic spectral response close to the normal-QG transition versus SR-QG transition lines. At the glass transition, $Z_{\delta}, \gamma_{\delta}$ and $\alpha_{\delta}$ vanish, which for the latter two results in zero energy excitations with infinite life-time and therefore infinite correlation times. The vanishing of the residuum $Z_{\delta}$ indicates that the discrete poles of the system, representing quasi-particles with weight $Z$, transform into a continuum represented by a branch cut in the complex plane as illustrated in Fig.~\ref{fig:PoleStruct}. As a consequence, a derivative expansion of the inverse propagator is no longer possible in the quantum glass phase.

When approaching the glass phase, the frequency interval which is described by classical relaxational dynamics (i.e. ${[0,\alpha_{\delta}]}$) shrinks and vanishes completely at the transition, where the system becomes quantum critical. The linear scaling of $\mathcal{A}(\omega)$ in combination with the closing of the spectral gap is taken in thermal equilibrium as the defining property of a quantum glass. However, for a general non-equilibrium setting, the closing of the spectral gap is only a necessary but not a sufficient condition for the glass phase. The unique property of the glass transition in a non-equilibrium setting is the emergence of a critical continuum at zero frequency, which leads to the closing of the gap of the retarded Green's function (distinct from the spectral gap).
From the structure of the low frequency response function, Eq.~\eqref{4Eq1}, we see that closing the spectral gap and a linear behavior of the spectral density is a non-trivial (and glass) signature only for a system where time-reversal symmetry is preserved, i.e. $\gamma=0$. On the other hand, the spectral gap closing is always present for broken time-reversal symmetry.

Within the glass phase, it is again possible to separate two distinct frequency regimes delimited by a cross-over frequency
\eq{4Eq7}{
\omega_c=2\kappa\left(1+\frac{\omega_0^2}{\omega_0^2+\kappa^2}+\frac{\left(\omega_0^2+\kappa^2\right)^2}{\sqrt{K}\omega_z^2}\right)^{-1},}
which depends on all system and bath parameters. For ${\omega<\omega_c}$, the atomic spectral density is described by
\eq{4Eq8}{
\mathcal{A}(\omega)\underset{\omega<\omega_c}{=}\mbox{sgn}(\omega)\sqrt{\frac{2\kappa\left(\omega_0^2+\kappa^2\right)|\omega|}{K\omega_0^2}}.}
This unusual square root behavior of the spectral density in the glass phase, illustrated in Fig.~\ref{fig:QGSpec} and also reflected in the pole structure Fig.~\ref{fig:PoleStruct}, is a characteristic feature for glassy systems that are coupled to an environment \cite{Cugliandolo2002, Strack2012}. It has been discussed previously in the context of metallic glasses, where collective charges couple to a bath of mobile electrons \cite{Strack2012} or for spin glasses, where the spins couple to an external ohmic bath \cite{Cugliandolo2002}. For intermediate frequencies, ${\omega>\omega_c}$, the spectral density is linear, as it is known for the non-dissipative zero temperature case. In the limit ${\kappa\rightarrow0}$, $\omega_c$ is shifted to smaller and smaller frequencies, vanishing for ${\kappa=0}$.
The cross-over frequency $\omega_c$ sets a time-scale ${t_c=\frac{1}{\omega_c}}$, such that for times ${t<t_c}$ the system behaves as if it were isolated and one would observe the behavior of a $T=0$ quantum glass for (relative) time scales $t<t_c$ in experiments. On the other hand, the long time behavior, $t>t_c$, of the atoms is described by overdamped dynamics, resulting from the coupling of the photons to a Markovian bath. This is a strong signature of low frequency equilibration of the atomic and photonic subsystem, which happens exactly at the glass transition (see Sec.~\ref{sec:Thermalization}).

\subsection{Atom-photon thermalization}\label{sec:AtDist}

We now discuss thermalization properties. The presence of quenched disorder in our model leads to an effective quartic atom-atom interaction, shown in Eq.~\eqref{2Eq35}, which allows for an energy redistribution to different frequency regimes. 

\subsubsection{Atom distribution function}
In order to determine the atomic distribution function ${F\sub{at}(\omega)}$, we make use of the FDR (see Appendix~\ref{sec:MvsQ}, Eq.~\eqref{MD12}), which for the atoms described by a scalar degree of freedom simplifies to
\eq{4Extra8}{
Q^K(\omega)=F\sub{at}(\omega)\left(Q^R(\omega)-Q^A(\omega)\right).}
The atomic correlation function $Q^K$ is determined via Eq.~\eqref{2Eq48}. This equation contains the photonic Keldysh Green's function via ${\Lambda^K(\omega)}$, and it is therefore evident, that the atomic distribution function will depend on the distribution function of the bare photons. The bare photon distribution function ${f\sub{ph}(\omega)}$ is again defined via the FDR, reading
\eq{4Eq9}{
G_0^K(\omega)=f\sub{ph}(\omega)\left(G^R_0(\omega)-G^A_0(\omega)\right),}
with the bare photon response and correlation functions $G^{R/A/K}_0$. Decomposing ${f=f\sub{S}+f\sub{AS}}$ into a symmetric ${f\sub{S}(\omega)=f\sub{S}(-\omega)}$ and an anti-symmetric ${f\sub{AS}(\omega)=-f\sub{AS}(-\omega)}$ contribution allows us to rewrite ${\Lambda^K(\omega)}$ in Eq.~\eqref{2Eq48} as
\begin{widetext}
\begin{eqnarray}
2\Lambda^K(\omega)&=&G^K_0(\omega)+G^K_0(-\omega)\nonumber\\
&=&2f\sub{AS}(\omega)\left(\Lambda^R(\omega)-\Lambda^A(\omega)\right)+f\sub{S}(\omega)\left(g^R(\omega)+g^R(-\omega)-g^A(\omega)-g^A(-\omega)\right)\nonumber\\
&=&2\left(f\sub{AS}(\omega)+\frac{\omega^2+\kappa^2+\omega_0^2}{2\omega\omega_0}f\sub{S}(\omega)\right)\left(\Lambda^R(\omega)-\Lambda^A(\omega)\right).\label{4Eq10}
\end{eqnarray}
\end{widetext}
Inserting this result into the expression for the correlation function \eqref{2Eq48}, and making use of Eq.~\eqref{2Eq49} and its complex conjugate yields
\eq{4Eq11}{
Q^K\sub{reg}=\left(f\sub{AS}+\frac{\omega^2+\kappa^2+\omega_0^2}{2\omega\omega_0}f\sub{S}\right)\left(Q^R-Q^A\right)}
and thus identifies the atomic distribution function
\eq{4Eq12}{
F\sub{at}(\omega)=f\sub{AS}(\omega)+\frac{\omega^2+\kappa^2+\omega_0^2}{2\omega\omega_0}f\sub{S}(\omega).}
This very general expression for the atomic distribution function incorporates the two most important examples, either a coupling to a thermal or a Markovian bath. For the coupling to a heat bath, the bare photonic distribution function is fully anti-symmetric with
${f\sub{AS}(\omega)=\coth\left(\frac{\omega}{2T}\right), f\sub{S}(\omega)=0}$, which implies that the atoms will be distributed according to a thermal distribution as well and experience the same temperature $T$ as the photons. For the case of dissipative photons, the bare distribution function of the photons is fully symmetric, with ${f\sub{S}(\omega)=1, f\sub{AS}(\omega)=0}$. Therefore the atomic distribution function for this system is
\eq{4Eq13}{
F\sub{at}(\omega)=\frac{\omega^2+\kappa^2+\omega_0^2}{2\omega\omega_0}.}

For small frequencies ${\omega\ll \sqrt{\omega_0^2+\kappa^2}}$, the atomic distribution function diverges as ${F\sub{at}(\omega)\sim\frac{1}{\omega}}$. This is the same asymptotic low frequency behavior as for the thermal distribution function ${\coth\left(\frac{\omega}{2T}\right)\sim\frac{2T}{\omega}}$, such that for low frequencies, the system is effectively described by a thermal distribution with a non-zero low frequency effective temperature (LET)
\eq{4Eq14}{
T\sub{eff}=\lim_{\omega\rightarrow0}\frac{\omega F\sub{at}(\omega)}{2}=\frac{\omega_0^2+\kappa^2}{4\omega_0}.}

The atomic distribution $F\sub{at}$ and low-frequency effective temperature $T\sub{eff}$ in Eqs.~\eqref{4Eq13}, \eqref{4Eq14} is identical to the distribution function and LET of the photonic $x$-component, which is obtained by expressing the photonic action in terms of the $x$ and $p$ component, ${p=\frac{i}{\sqrt{2}}(\cre{a}{}-a)}$, and subsequently integrating out the $p$ component. This procedure is shown in App. \ref{sec:XDist}. From the resulting action, the $x$ component is described by a distribution function ${F_{xx}(\omega)=F\sub{at}(\omega)}$, resulting from the coupling of the photons to the Markovian bath. Due to the strong atom-photon interaction, the atoms and the photonic $x$ component equilibrate, resulting in the same distribution function and LET.

\subsubsection{Photon distribution function}\label{sec:Thermalization}

To compute the photon distribution function, we use the FDR 
\eq{4Eq28}{
G^K(\omega)=G^R(\omega) F\sub{ph}(\omega)-F\sub{ph}G^A(\omega),}
which in this case is an equation for the $2\times2$ matrices $G^{R/A/K}$ and $F$. The matrix $F$ solving Eq.~\eqref{4Eq28} is not diagonal, and the distribution of the excitations is determined by its eigenvalues $f_{\alpha}$. These are shown in Fig.~\ref{fig:Thermalization} and illustrate the thermalization process of the system. In the normal and superradiant phase, the photons have a lower LET than the atoms, resulting from the frequency regime for which the dynamics is classical relaxational. As for the spectral response, when the glass transition is approached, this classical region is shifted towards $\omega=0$ and the photon LET approaches the atomic effective temperature. At the transition, the photons and atoms have thermalized completely in the low frequency regime.

\subsection{Emergent photon glass}\label{sec:PhotonGlass}

In the glass phase, the condensate order parameter $\langle a_i\rangle\propto\frac{1}{N}\sum_l\langle\sigma_l^x\rangle=\psi$ vanishes for all photon modes $(i)$. However there exists a photon version of the Edwards-Anderson parameter 
\eq{4Eq23}{
\tilde{q}\sub{EA}=\lim_{\tau\rightarrow\infty} \frac{1}{M}\sum_{i=1}^M\langle x_i(t+\tau) x_i(t)\rangle\propto q\sub{EA},}
where $x=\frac{1}{\sqrt{2}}(a+a^{\dagger})$ is the photon $x$ operator and Eq.~\eqref{4Eq23} only holds for the $x$-$x$ correlations (and for those with finite contributions to $x$-$x$). A non-vanishing photon Edwards-Anderson parameter implies an infinite correlation time for the photons, analogous to the atomic $q\sub{EA}$-parameter. This is best illustrated by the correlator in the complex basis
\eq{4Extra9}{
\lim_{\tau\rightarrow\infty}\langle a(t+\tau)a^{\dagger}(t)\rangle=\frac{1}{2}\lim_{\tau\rightarrow\infty}\langle x(t+\tau)x(t)\rangle=\frac{\tilde{q}\sub{EA}}{2},}
where we made use of the fact that the $x$-$p$ and $p$-$p$ correlations vanish for ${\tau\rightarrow\infty}$. Eq.~\eqref{4Extra9} implies that a photon which enters the cavity at time $t$ has a non-vanishing probability to decay from the cavity at arbitrary time $t+\tau$, with $\tau\in[0,\infty)$. This highlights a connection to photon localization in disordered media \cite{Angelani2006,Andreasen2011}.

Close to the glass transition, the properties of the atomic system are completely mapped to the inverse photon Green's function.
In the low frequency and small $\kappa$ limit, i.e. $\omega,\kappa\ll\omega_0,\omega_z$, the inverse photon Green's function Eq.~(\ref{2Eq57}) has the expansion
\eq{4Eq26}{
D_{x-p}^R(\omega)=\left(\begin{array}{cc} -\frac{\omega_0\omega_zD^R\sub{at}(\omega)}{2\left(\omega^2-\lambda\right)} &0\\ 0 &-\omega_0\end{array}\right),}
such that the atomic low frequency physics is mapped to the photon $x$-$x$ component.

\begin{figure}[t!]
\includegraphics[width=1\linewidth]{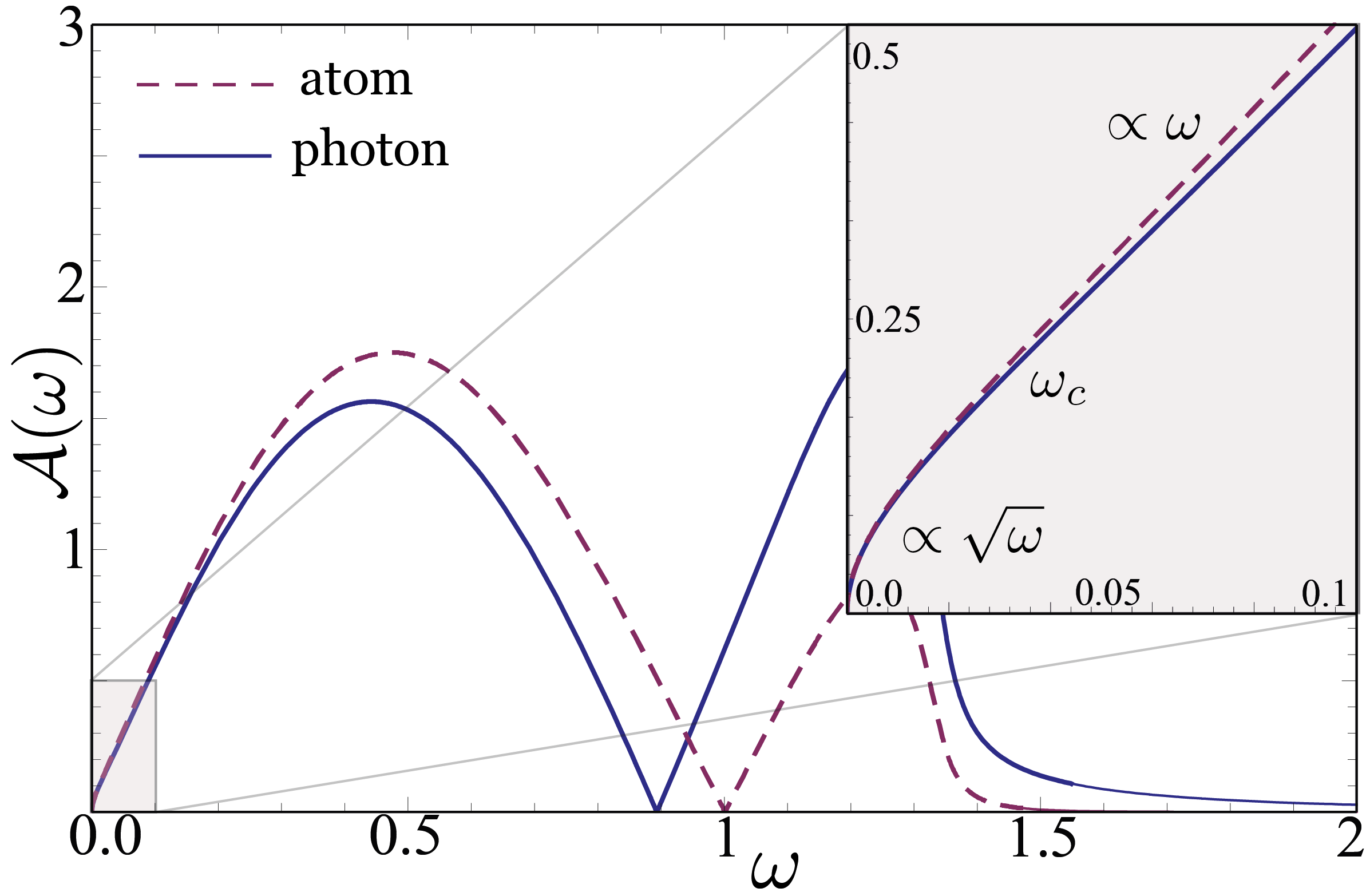}
    \caption
    {\label{fig:QGSpecPhoton}
    (Color online) \emph{Spectral equilibration:} Photon $x$-$x$ spectral response $\mathcal{A}_{xx}(\omega)=-2\mbox{Im}\left(G^R_{xx}(\omega)\right)$ in the glass phase for parameters $K=0.04, J=0.12, \omega_z=2, \omega_0=1, \kappa=0.02$. As for the SR phase, it shows the same low frequency behavior as the atomic spectral response $-2\mbox{Im}\left(Q^R(\omega)\right)$ (multiplied with a constant $\frac{\omega_z\omega_0}{2\lambda}$). As for the atomic spectral density, one can clearly identify the overdamped regime with the unusual square-root behavior and the linear regime, separated by the frequency $\omega_c$.}
\end{figure}

The determinant of $D^R_{x-p}$ vanishes at the zeros of $D^R\sub{at}$, such that the photon propagator shows the same poles or branch cuts as the atomic propagator, and the scaling behavior at the glass transition obtained from the photons is identical to the one obtained from the atoms. The photon response properties induced by the atom-photon coupling are most pronounced in the $x$-$x$ component $G^R_{xx}$ of the retarded photon Green's function, 
\eq{4Eq27}{G^R_{xx}(\omega)=\frac{\omega_0}{\left(\omega+i\kappa\right)^2+2\omega_0\Sigma^R(\omega)-\omega_0^2}.}
For low frequencies, we can perform the same approximation as above to find
\eq{4Eq27A}{G^R_{xx}(\omega)=\frac{2\left(\lambda-\omega^2\right)}{\omega_0\omega_zD^R\sub{at}(\omega)}=\frac{2\left(\lambda-\omega^2\right)}{\omega_0\omega_z}Q^R(\omega).}

Close to the glass transition and in the glass phase, the atomic retarded Green's function $Q^R$ determines the low frequency photon $x$-$x$ response function. 
This is reflected in Fig.~\ref{fig:QGSpecPhoton}.
The discussion of the atomic response and scaling behavior in Sec.~\ref{sec:AtSpec} remains valid for the photons. 

\subsection{Cavity glass microscope}\label{sec:Exp}

We now describe three experimental signatures (cavity output fluorescence spectrum, photon real-time correlation function $g^{(2)}(\tau)$, and the photon response via driven homodyne detection) of the superradiant and glassy phases and their spectral properties. The cavity output is determined by the cavity input and the intra-cavity photons via the input-output relation \cite{Gardiner1985, Collett1984}
\eq{5Eq2}{
a^{\phantom{\dagger}}\sub{out}(\omega)=\sqrt{2\kappa}\ \tilde{a}(\omega)+a^{\phantom{\dagger}}\sub{in}(\omega),}
with the cavity input annihilation operator $a^{\phantom{\dagger}}\sub{in}(\omega)$ and the averaged intra-cavity field
\eq{5Eq3}{
\tilde{a}(\omega)=\frac{1}{\sqrt{M}}\sum_{i=1}^M \ann{a}{i}(\omega)}
accounting for the $M$ distinct cavity modes. 

\subsubsection{Cavity output fluorescence spectrum}
The fluorescence spectrum $S(\omega)$ describes the (unnormalized) probability of measuring a photon of frequency $\omega$ at the cavity output \cite{Dimer2007}, and is defined by
\eq{5Eq1}{
S(\omega)=\langle \cre{a}{\mbox{\tiny out}}(\omega)\ann{a}{\mbox{\tiny out}}(\omega)\rangle,}
where $\cre{a}{\mbox{\tiny out}}(\omega),\ann{a}{\mbox{\tiny out}}(\omega)$ are creation, annihilation operators of the output field. Considering a vacuum input field, the fluorescence spectrum is expressed solely by the auto-correlation function of the intra-cavity field
\eq{5Eq4}{
S(\omega)=\langle \tilde{a}^{\dagger}(\omega)\tilde{a}(\omega)\rangle=\int_{\tau} e^{i\omega\tau}\langle\tilde{a}^{\dagger}(0)\tilde{a}(\tau)\rangle=iG^{<}(\omega),}
which is the ``$G$-\textit{lesser}'' Green's function, occurring in the $(\pm)$-representation (see  \cite{KamenevBook, AltlandBook}). Introducing also the ``$G$-\textit{greater}'' Green's function
\eq{5Extra1}{
iG^>(\omega)=\int_{\tau} e^{i\omega\tau}\langle\tilde{a}(\tau)\tilde{a}^{\dagger}(0)\rangle,}
we can express response and correlation functions in terms of  $G^{</>}$
\begin{eqnarray}
G^{K}(\omega)&=&G^{>}(\omega)+G^{<}(\omega)\label{5Eq5}\\
G^R(\omega)-G^A(\omega)&=&G^>(\omega)-G^<(\omega),\label{5Eq6}
\end{eqnarray}
which yields
\begin{eqnarray}
S(\omega)&=&\frac{i}{2}\left(G^K(\omega)-G^R(\omega)+G^A(\omega)\right)\nonumber\\
&=&\frac{i}{2}\left( G^R(\omega)\left(F(\omega)-1\right)-\left(F(\omega)-1\right)G^A(\omega)\right).\ \ \ \ \ \ \ \label{5Eq7}
\end{eqnarray}
In thermal equilibrium ($F(\omega)=2n_B(\omega)+1$), where $F$ is diagonal in Nambu space, this expression simply reads
\eq{5Extra2}{
S(\omega)=n_B(\omega)\mathcal{A}(\omega)}
and the fluorescence spectrum reveals information about the intra-cavity spectral density $\mathcal{A}$.

In order to further analyze Eq.~\eqref{5Eq7}, we decompose the fluorescence spectrum into a regular part and a singular part as it was done for the Keldysh Green's function in Eq.~\eqref{2Eq42},
\eq{5Eq8}{
S(\omega)=S\sub{reg}(\omega)+2\pi \tilde{q}\sub{EA}\delta(\omega),}
with the Edwards-Anderson parameter for the photons $\tilde{q}\sub{EA}$. 
The regular part $S\sub{reg}$ is determined by the regular contributions from the response and distribution function, $G^{R/A}, F$, which we have analyzed in the previous section. For small frequencies, $F(\omega)\propto\frac{1}{\omega}$ and in the normal and SR phase, $(G^R-G^A)\propto\omega$, which leads to a finite contribution of $S\sub{reg}$ to the spectrum. In contrast, in the QG phase, $(G^R-G^A)\propto\sqrt{\omega}$, such that
\eq{5Extra3}{
S\sub{reg}(\omega)\propto\frac{1}{\sqrt{\omega}}}
has a square root divergence for small frequencies $\omega<\omega\sub{QG}^c$ (see Eq.~\eqref{4Eq7}). This divergence is indicated in Fig.~\ref{fig:Fluorescence}~(c) and is a clear experimental signature of the glass phase.

A further distinction between all three phases is possible by decomposing the fluorescence spectrum into a coherent and an incoherent part, where the coherent part describes the ``classical'' solution (i.e. the part resulting from the presence of a photon condensate $\langle\tilde{a}\rangle\neq0$) and the incoherent part describes the fluctuations. Accordingly, the coherent part is
\eq{5Eq9}{
S\sub{c}(\omega)=2\pi |\langle\tilde{a}\rangle|^2\delta(\omega)}
and the incoherent part reads
\eq{5Eq10}{
S\sub{inc}(\omega)=S\sub{reg}(\omega)+2\pi\left( \tilde{q}\sub{EA}-|\langle\tilde{a}\rangle|^2\right)\delta(\omega).}
Typical fluorescence spectra characterizing the three distinct phases are plotted in Fig.~\ref{fig:Fluorescence}. For the normal phase, the spectrum shows central and outer doublets associated with the hybridized atomic and photonic modes. Above the critical point for the superradiance transition, the doublets merge since a single mode becomes critical. However, compared to the single-mode transition, the central peak is much broader as a consequence of disorder. Additionally, in the superradiant phase, the fluorescence spectrum has a non-zero coherent contribution, which allows for a unique identification of this phase. 

In the glass phase, the doublets have merged after the emergence of a critical continuum of modes at $\omega=0$, and one can clearly identify the square root divergence for small frequencies, as discussed above. Additionally, the singular behavior of $S(\omega)$ in the glass phase is of incoherent nature, since $\langle\cre{a}{}\rangle=0$. This combination of an incoherent zero frequency peak together with the absence of a coherent contribution uniquely defines the fluorescence spectrum in the glass phase and allows for a complete classification of the system's phases via fluorescence spectroscopy.

The coherent contribution to the spectrum can be determined via homodyne detection (see below), where $\langle\tilde{a}\rangle$ can be measured directly.

\subsubsection{Photon real-time correlation function $g^{(2)}(\tau)$}

The time-resolved four-point correlation function of the output field
\eq{g2Eq1}{
g^{(2)}(t,\tau)=\frac{\langle \cre{a}{\mbox{\tiny out}}(t)\cre{a}{\mbox{\tiny out}}(t+\tau)\ann{a}{\mbox{\tiny out}}(t+\tau)\ann{a}{\mbox{\tiny out}}(t)\rangle}{|\langle \cre{a}{\mbox{\tiny out}}(t)\ann{a}{\mbox{\tiny out}}(t)\rangle|}}
reveals how the correlations in the cavity decay with the time difference $\tau$. In steady state, $g^{(2)}(t,\tau)$ only depends on the time difference $\tau$ and we write $g^{(2)}(\tau)$. For $\tau\rightarrow0$, $g^{(2)}(0)$ is a measure of the underlying photon statistics in the cavity, e.g. indicates bunching or anti-bunching of the cavity photons, respectively. 

In the open Dicke model, due to the effective temperature (cf. Fig.~\ref{fig:Thermalization} and Ref.~\cite{Dalla2012}),  $g^{(2)}(0)>1$, describing photon bunching, as expected for thermal bosons. We find $g^{(2)}(0)=3$ for all the three phases, which stems from the off-diagonal atom-photon coupling in the Dicke model and coincides with the findings in Ref.~\cite{oztop2012} for the normal and superradiant phase.

In the normal and superradiant phase, the long time behavior is governed by the classical low frequency dynamics, leading to an exponential decay
\eq{g2Eq2}{
g^{(2)}(\tau)\sim 1+2e^{-2\kappa\tau}.}
This behavior is well known for the single mode Dicke model \cite{oztop2012} and remains valid for the multimode case, away from the glass transition. In contrast, when the glass phase is approached, the modes of the system form a branch cut in the complex plane and the correlation function in the glass phase decays algebraically, according to
\eq{g2Eq3}{
g^{(2)}(\tau)\sim 1+\left(\frac{\tau_0}{\tau}\right)^{\frac{1}{2}},}
where $\tau_0=O(1/\omega_0)$. This algebraic decay of the correlation function provides clearcut evidence for a critical continuum of modes around zero frequency witnessing the glass phase. In Fig.~\ref{fig:g2}, we show $g^{(2)}(\tau)$ demonstrating this behavior.

In order to compute the four-point correlation function \eqref{g2Eq1}, we make use of Eq.~\eqref{5Eq2} and the vacuum nature of the input field, i.e. the fact that all averages over $a\sub{in}, a^{\dagger}\sub{in}$ vanish.
As a consequence, the operators for the output field in Eq.~\eqref{g2Eq1} can be replaced by the operators for the averaged cavity field $\tilde{a}$, see Eq.~\eqref{5Eq3}. The denominator in Eq.~\eqref{g2Eq1} is then
\eq{g2Eq4}{|\langle \tilde{a}^{\dagger}(t)\tilde{a}(t)\rangle|^2=|\langle \crea{\tilde{a}}{-}(t)\ann{\tilde{a}}{+}(t)\rangle|^2=|G^<(0)|^2.}
The numerator similarly is expressed as
\eq{g2Eq5}{
\langle\tilde{a}^{\dagger}(t)\tilde{a}^{\dagger}(t+\tau)\tilde{a}(t+\tau)\tilde{a}(t)\rangle=\langle\crea{\tilde{a}}{-}(t)\crea{\tilde{a}}{-}(t+\tau)\ann{\tilde{a}}{+}(t+\tau)\ann{\tilde{a}}{+}(t)\rangle .}
Note that both expressions (Eqs.~\eqref{g2Eq4}, \eqref{g2Eq5}) preserve the correct operator ordering of Eq.~\eqref{g2Eq1}, according to the different time-ordering on the $(+)$, $(-)$-contour, respectively.

The four-point function in Eq.~\eqref{g2Eq5} can be expressed in terms of functional derivatives of the partition function $\mathcal{Z}$ (Eq.~\eqref{2Eq49}) with respect to the source fields $\mu$ (Eq.~\eqref{2Eq29}). In the thermodynamic limit, the macroscopic action, Eq.~\eqref{2Eq38}, depends only on atomic and photonic two-point functions and, equivalent to Wick's theorem, the four-point function becomes the sum over all possible products of two-point functions
\begin{eqnarray}
G^{(2)}(\tau)&=&\langle\crea{\tilde{a}}{-}(t)\crea{\tilde{a}}{-}(t+\tau)\ann{\tilde{a}}{+}(t+\tau)\ann{\tilde{a}}{+}(t)\rangle\nonumber\\
&=&\langle\crea{\tilde{a}}{-}(t)\ann{\tilde{a}}{+}(t)\rangle\ \langle\crea{\tilde{a}}{-}(t+\tau)\ann{\tilde{a}}{+}(t+\tau)\rangle\nonumber\\
& &+\langle\crea{\tilde{a}}{-}(t+\tau)\ann{\tilde{a}}{+}(t)\rangle\ \langle\crea{\tilde{a}}{-}(t)\ann{\tilde{a}}{+}(t+\tau)\rangle\nonumber\\
& &+\langle\crea{\tilde{a}}{-}(t)\crea{\tilde{a}}{-}(t+\tau)\rangle\ \langle\ann{\tilde{a}}{+}(t+\tau)\ann{\tilde{a}}{+}(t)\rangle\nonumber\\
&=&|G^<(0)|^2+|G^<(\tau)|^2+|G^<\sub{an}(\tau)|^2,\label{g2Eq6}
\end{eqnarray}
with the anomalous $G$-\textit{lesser} function
\eq{g2Eq7}{
G^<\sub{an}(\tau)=-i\langle \tilde{a}(\tau)\tilde{a}(0)\rangle.}
Inserting Eq.~\eqref{g2Eq6} into the expression for the four-point correlation function yields
\eq{g2Eq8}{
g^{(2)}(\tau)=1+|g^{(1)}(\tau)|^2+\frac{|G^<\sub{an}(\tau)|^2}{|G^<(0)|^2},}
with the two-point correlation function
\eq{g2Eq9}{
g^{(1)}(\tau)=\frac{G^<(\tau)}{G^<(0)}=\frac{\langle \tilde{a}^{\dagger}(\tau)\tilde{a}(0)\rangle}{\langle \tilde{a}^{\dagger}(0)\tilde{a}(0)\rangle}.}

$G^<(\tau)$ is the Fourier transform of the fluorescence spectrum $S(\omega)$, as discussed in the previous section, and we therefore decompose it according to
\eq{g2Extra1}{
G^<(\tau)=\tilde{q}\sub{EA}+G^<\sub{reg}(\tau),}
with $G^<\sub{reg}(\tau)$ being the Fourier transform of $S\sub{reg}(\omega)$.
In the infinite time limit, the regular part of $G^<(\tau)$ decays to zero, such that the infinite correlation time value becomes
\eq{g2Extra2}{
g^{(1)}(\tau)\underset{\tau\rightarrow\infty}{=}\frac{\tilde{q}\sub{EA}}{\tilde{q}\sub{EA}+G^<\sub{reg}(0)}=\frac{\tilde{q}\sub{EA}}{\tilde{q}\sub{EA}+N\sub{reg}}.}
$N\sub{reg}$ denotes the occupation of the non-critical cavity modes. The way how this  value of $g^{(1)}$ is reached in time is determined by the $\frac{1}{\sqrt{\omega}}$ divergence of $S\sub{reg}(\omega)$ for frequencies $\omega<\omega_c$ smaller than the cross-over frequency $\omega_c$, cf. Fig.~\ref{fig:Fluorescence}. This leads to
\eq{g2Eq10}{
g^{(1)}(\tau)\underset{\tau>\tau_c}{\rightarrow}\frac{\tilde{q}\sub{EA}}{\tilde{q}\sub{EA}+N\sub{reg}}+\left(\frac{\tilde{\tau}_0}{\tau}\right)^{\frac{1}{2}},}
where $\tau_c=\frac{2\pi}{\omega_c}$ and $\tilde{\tau}_0$ has to be determined numerically. This algebraic decay to the infinite $\tau$ value of the correlation function with the exponent $\nu=\frac{1}{2}$ has also been found in Ref.~\cite{Cugliandolo1999} for the correlation function of a spin glass coupling to a finite temperature ohmic bath, in line with the discussion of universality in Sec. \ref{sec:AtSpec}. The finite temperature exponent results from the non-zero effective temperature of the system, which influences the correlation function. For the case of  $T\sub{eff}=0$ this exponent changes to $\nu=\frac{3}{2}$ but the spectral properties are left unchanged.

The non-zero value of the two-point correlation $g^{(1)}(\tau)\rightarrow\frac{\tilde{q}\sub{EA}}{\tilde{q}\sub{EA}+N\sub{reg}}$ for $\tau\rightarrow\infty$ serves as a possible measure of the photonic Edwards-Anderson parameter $\tilde{q}\sub{EA}$ in the glass phase: $\tilde{q}\sub{EA}$ can be inferred from a correlation measurement, if the total photon number in the cavity $N\sub{tot}=\tilde{q}\sub{EA}+N\sub{reg}$ has been measured separately.

Taking the absolute value of $g^{(1)}(\tau)$ in Eq.~\eqref{g2Eq10}, leads to the dominant contribution
\eq{g2Eq11}{
|g^{(1)}(\tau)|^2\underset{\tau>\tau_c}\rightarrow\left( \frac{\tilde{q}\sub{EA}}{\tilde{q}\sub{EA}+N\sub{reg}}\right)^2+2\frac{\tilde{q}\sub{EA}}{\tilde{q}\sub{EA}+N\sub{reg}}\left(\frac{\tilde{\tau}_0}{\tau}\right)^{\frac{1}{2}},}
as displayed in the asymptotic behavior of the four-point correlation function (Eq.~\eqref{g2Eq3} , where we have absorbed the prefactors in the definition of $\tau_0$ and normalized the long-time limit to unity.

While the non-zero  value of $g^{(1)}(\tau \to \infty)$ is caused by critical poles of the system, it does not include any more information about the pole structure of the system and may for instance be caused by a single critical pole, as it is the case for the superradiance transition. However, the algebraic decay to the infinite correlation time value of $g^{(1)}$, and the same for $g^{(2)}$, is a clear signature of a branch cut in the complex plane and therefore a continuum of modes reaching to zero frequency. This in turn is a strong signature of the critical glass phase in the cavity.

\subsubsection{Photon response via driven homodyne detection}
Here we relate homodyne detection measurements of the output signal to the quadrature response functions in the Keldysh formalism and calculate the corresponding signal. This gives predictions for the experimental analysis of the spectral properties and the scaling at the glass transition, which have been discussed in previous sections.

In the process of homodyne detection, the output field $a\sub{out}$ is sent to a beam-splitter, where it is superimposed with a coherent light field $\beta(t)=\beta\ e^{-i(\omega_{\beta}t+\theta)}$ with frequency $\omega_{\beta}$, amplitude $\beta$ and phase $\theta$. After passing the beam-splitter, the intensity of the two resulting light fields is measured and the difference in this measurement (the difference current) for the case of a $50/50$ beam-splitter is described by
\begin{eqnarray}
n_{-}(t)&=&i\left\langle a^{\dagger}\sub{out}(t)\beta(t)-\beta^{*}(t)a^{\phantom{\dagger}}\sub{out}(t)\right\rangle\nonumber\\
&=&\beta\left\langle e^{i(\theta-\frac{\pi}{2})} a\sub{out}(t)e^{i\omega_{\beta}t}+e^{-i(\theta-\frac{\pi}{2})}a\sub{out}^{\dagger}(t)e^{-i\omega_{\beta}t}\right\rangle.\label{5Eq12}
\end{eqnarray}
Here, we added a conventional phase shift $\phi=\frac{\pi}{2}$ of the beam-splitter. For the case of a vacuum input field, Eq.~\eqref{5Eq12} simply measures the steady state expectation value of the cavity quadrature components 
\eq{5Eq13}{
X_{\theta-\frac{\pi}{2},\omega_{\beta}}(t)=e^{i(\theta-\frac{\pi}{2})} \tilde{a}(t)e^{i\omega_{\beta}t}+e^{-i(\theta-\frac{\pi}{2})}\tilde{a}^{\dagger}(t)e^{-i\omega_{\beta}t},}
with the intra-cavity operators $\tilde{a}$, as defined in Eqs.~\eqref{5Eq2}, \eqref{5Eq3}. This quantity indicates a finite superradiance order parameter $\langle\tilde{a}\rangle$, but for the steady state contains no further information.

\begin{figure}[t]
\vspace{1mm}
\includegraphics[width=1.\linewidth]{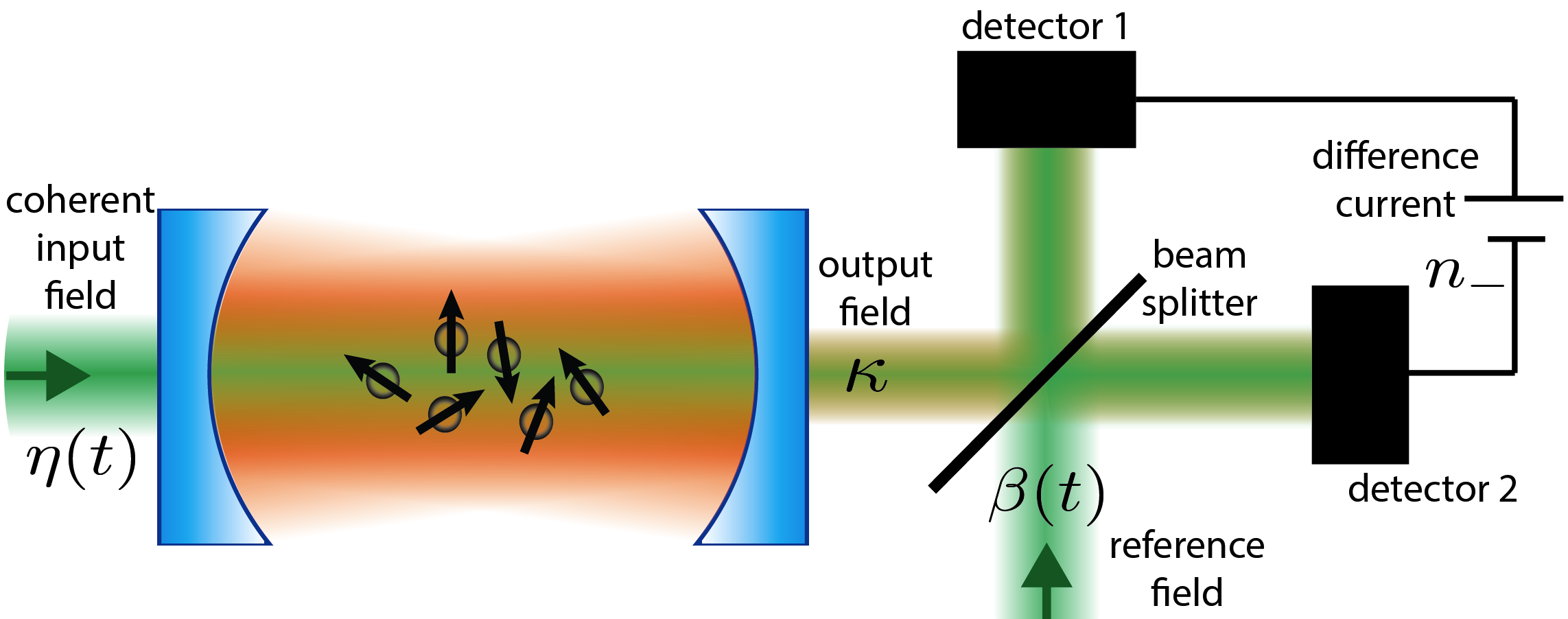}
    \caption
    {\label{fig:Homodyne}
    (Color online) Illustration of homodyne detection of a weakly driven cavity. The cavity is driven via a weak coherent input field $\eta(t)$ entering the cavity through one of the mirrors. Then a homodyne measurement is performed on the output signal of the driven cavity. For this, the output signal is superimposed with a reference laser $\beta(t)$ via a $50/50$ beam-splitter and the difference current of the two outgoing channels is measured. From this, the response function of the photons in the cavity can be measured by tuning the relative phases and frequencies of $\beta(t)$ and $\eta(t)$, as explained in the text.}
\end{figure}

 This situation changes when the input field is changed from the vacuum state to a weak coherent laser field $\eta(t)$. For this special case, the difference current in Eq.~\eqref{5Eq12} is modified according to
\begin{eqnarray}
n_{-}(t)&=&i\left(\eta^{*}\beta-\beta^{*}\eta\right)(t)+i\sqrt{2\kappa}\left\langle \tilde{a}^{\dagger}(t)\beta(t)-\beta^{*}(t)\tilde{a}(t)\right\rangle\nonumber\\
&=&i\left(\eta^{*}\beta-\beta^{*}\eta\right)(t)+\sqrt{2\kappa}|\beta|\left\langle X_{\theta-\frac{\pi}{2},\omega_{\beta}}(t)\right\rangle.\label{5Eq14}
\end{eqnarray}
For the special case of the input field coming from the same signal as the reference laser, we have $\eta(t)=\beta(t)$ (which we assume from now on for simplicity) and the first term in Eq.~\eqref{5Eq14} vanishes. The main difference here, is that the quadrature operator $X_{\theta,\omega_{\beta}}(t)$ is not evaluated for the steady state but for a state which has been perturbed by the weak laser field $\beta(t)$. For a weak laser amplitude $|\beta|\ll 1$, the system stays in the linear response regime and the difference current is proportional to the retarded Green's function for the quadrature component $X_{\theta+\frac{\pi}{2},\omega_{\beta}}$ as we proceed to show. \\
The interaction between the cavity photons and the radiation field outside the cavity is commonly described by the Hamiltonian
\eq{5Eq15}{
H\sub{int}=i\sqrt{2\kappa}\left(\tilde{a}^{\dagger}a^{\phantom{\dagger}}\sub{in}-a^{\dagger}\sub{in}\tilde{a}\right),}
which after a transformation to the Keldysh action and replacing the input fields by the coherent light field $\beta(t)$ enters the action as
\eq{5Eq16}{
S\sub{int}=\sqrt{2\kappa}\int_{\omega} \left(\tilde{a}_{c}^{*},\tilde{a}_{q}^{*}\right)(\omega)i\sigma^x\left(\begin{array}{c}\beta_{c}\\ \beta_{q}  \end{array}\right)(\omega)+\mbox{h.c.},}
which is exactly the form of a source term in quantum field theory, generating all Green's functions of the system via functional derivatives with respect to the fields $\beta$. Expressing the action \eqref{5Eq16} in terms of Keldysh components of the quadrature fields $X_{\theta,\omega_{\beta}}$ yields
\eq{5Eq17}{
S\sub{int}=\sqrt{2\kappa}\int_{\omega}\left(X_{c,\theta-\frac{\pi}{2},\omega_{\beta}},X_{q,\theta-\frac{\pi}{2},\omega_{\beta}}\right)(\omega)\ \sigma^x\left(\begin{array}{c}|\beta_{c}|\\ |\beta_{q}|  \end{array}\right)(\omega).}
The linear response $\langle X_{\theta-\frac{\pi}{2},\omega_{\beta}}\rangle^{(1)}(t)$ of the quadrature expectation value is expressed as (see appendix \ref{sec:LinResp})
\eq{5Eq18}{
\langle X_{\theta-\frac{\pi}{2},\omega_{\beta}}\rangle^{(1)}(t)=-2\kappa|\beta|^2\int_{t'}G^R_{X_{\theta-\frac{\pi}{2}},\omega_{\beta}}(t-t'),}
with the quadrature response function
\eq{5Eq19}{
G^R_{X_{\theta-\frac{\pi}{2}},\omega_{\beta}}(t-t')=-i\theta(t-t')\langle\left[X_{\theta-\frac{\pi}{2},\omega_{\beta}}(t),X_{\theta-\frac{\pi}{2},\omega_{\beta}}(t') \right]\rangle.}
For the specific choice of $\theta=\frac{\pi}{2}$, 
\eq{ExtraHom}{
X_{\theta-\frac{\pi}{2},\omega_{\beta}}=\sqrt{2}x_{\omega_{\beta}}=\left( \tilde{a}(t)e^{i\omega_{\beta}t}+\tilde{a}^{\dagger}(t)e^{-i\omega_{\beta}t}\right),}
the response function $G^R_{X_{\theta-\frac{\pi}{2}},\omega_{\beta}}(t-t')=2G^R_{x_{\omega_{\beta}}}(t-t')$ becomes the $x$-$x$ retarded Green's function in a frame rotating with the laser frequency $\omega_{\beta}$. In this case, the difference current is a direct measurement of the $x$-$x$ response
\eq{5Eq20}{
n_-(t)=-4\kappa |\beta|^2\int_{t'} G^R_{x_{\omega_{\beta}}}(t-t'),}
which we have discussed in detail in Sec.~\ref{sec:PhotonGlass}. The frequency dependence of $x_{\omega_{\beta}}$, indicated by the subscript $\omega_{\beta}$, coming from Eq.~\eqref{5Eq13}, can be used to scan through different frequency regimes and directly access the atom and photon $x$-$x$ spectral response.

\section{Conclusion}

We have developed the non-equilibrium theory of the multimode Dicke model with quenched disorder and Markovian dissipation, and provided a comprehensive characterization of the resulting phases in terms of standard experimental observables. The main theoretical findings relate to the interplay of disorder and dissipation. We establish the robustness of a disorder induced glass in the presence of Markovian dissipation. This concerns, for example, the presence of an Edwards-Anderson order parameter and the algebraic decay of correlation functions in the entire glass phase. Central quantitative aspects, such as the decay exponents of the correlation functions, are strongly affected by the presence of dissipation. Disorder leads to enhanced equilibration of the atomic and photonic subsystems for both the spectral (response) and their statistical properties. The spin glass physics of the atoms is mirrored onto the photonic degrees of freedom. We presented direct experimental signatures for the atomic and photonic dynamics that allow unambiguous characterization of the various superradiant and glassy phases.

Several directions for future work emerge from these results. In particular, the realization of disorder may not be governed by an ideal single Gaussian probability distribution in experimental realizations of multimode Dicke models. This may concern, for example, effects relating to the finite number of cavity modes ($M$) or effective two-level atoms ($N$). $1/N$ corrections contain information on the critical behavior close to the conventional Dicke transition \cite{nagy11,Dalla2012}, with similar features expected for the glass transition. While we expect the main glassy features to be robust to such finite-size effects, it would be interesting to study a concrete cavity geometry with specific information of the cavity mode functions.

Furthermore, with our focus on the stationary state we did not touch upon the interesting questions of glassy dynamics \cite{Kennett2001,Cugliandolo2002} in this work (for thermalization dynamics of the single mode Dicke model, see \cite{PhysRevLett.108.073601}). An interesting problem is a quantum quench of the open, disordered system. In particular, the non-universal short time and transient regimes should contain more system specific and non-equilibrium information. In the long time limit, the nature of aging and dependencies on the aging protocol remains to be explored.

\begin{acknowledgments}

The authors thank F. Brennecke, E. Dalla Torre, T. Donner, T. Esslinger, S. Gopalakrishnan, J. Leonard, B. L. Lev, M. D. Lukin, L. M. Sieberer, and P. Zoller for helpful discussions, and F. Brennecke, T. Donner and S. Gopalakrishnan for a critical reading of the manuscript. This work was supported by the DFG under grant Str 1176/1-1, by the NSF under Grant DMR-1103860, by the Army Research Office Award W911NF-12-1-0227, by the Center for Ultracold Atoms (CUA) and by the Multidisciplinary University Research Initiative (MURI) (P.S.), the Austrian Science Fund (FWF) through SFB FOQUS F4016-N16 and the START grant Y 581-N16 (S. D.), the European Commission (AQUTE, NAMEQUAM), the Institut f\"ur Quanteninformation GmbH and the DARPA OLE program.

\end{acknowledgments}

\appendix

\section{Photon fields for superradiant phase}\label{app:Nambu}
In order to describe a system where the particle number is not conserved, as it is the case for the photons in the Dicke model, we introduce the spinor field
\eq{Nambu1}{
\ann{A}{\alpha,j}(t)=\left(\begin{array}{c} \ann{a}{\alpha,j}(t)\\ \crea{a}{\alpha,j}(t)\end{array}\right),}
containing the bosonic fields $\ann{a}{\alpha,j}(t), \crea{a}{\alpha,j}(t)$ for a quantum state $j$ and with index $\alpha=q,c$. The corresponding adjoint field is
\eq{Nambu2}{
\cre{A}{\alpha,j}(t)=\left(\crea{a}{\alpha,j}(t), \ann{a}{\alpha,j}(t)\right).}
The action for a quadratic problem is (for simplicy we consider only a single quantum state)
\eq{Nambu3}{
S=\int_{t,t'} \left( \cre{A}{c}(t), \cre{A}{q}(t)\right)D_{4\times4}(t,t') \left(\begin{array}{c} A_{c}(t')\\ \ann{A}{q}(t')\end{array}\right),}
where
\eq{Nambu4}{D_{4\times4}(t,t')=\left(\begin{array}{cc} 0& D^A_{2\times2}(t,t')\\ D^R_{2\times2}(t,t') & D^K_{2\times2}(t,t')\end{array}\right)=\left(G_{4\times4}\right)^{-1}(t,t')}
is the inverse Green's function. The Keldysh correlation and retarded Green's function are also $2\times2$ matrices, which can be expressed in terms of operator averages according to
\begin{eqnarray}
G^R_{2\times2}(t,t')&=&\left(D^R_{2\times2}\right)^{-1}(t,t')\label{Nambu5}\\
&=&-i\theta(t-t')\left\langle\left(\begin{array}{cc}
[a(t),a^{\dagger}(t')] & [a(t),a(t')] \\ \phantom{.}[a^{\dagger}(t),a^{\dagger}(t')] &  [a^{\dagger}(t),a(t')] \end{array}\right) \right\rangle\nonumber\end{eqnarray}
and
\begin{eqnarray}
G^K_{2\times2}(t,t')&=&-\left(G^R_{2\times2}\circ D^K_{2\times2}\circ G^A_{2\times2}\right)(t,t')\label{Nambu6}\\
&=&-i\left\langle\left(\begin{array}{cc}
\{a(t),a^{\dagger}(t')\} & \{a(t),a(t')\} \\ \phantom{.}\{a^{\dagger}(t),a^{\dagger}(t')\} &  \{a^{\dagger}(t),a(t')\} \end{array} \right)\right\rangle\nonumber.\end{eqnarray}
In Eq.~\eqref{Nambu6}, the $\circ$-operation represents convolution with respect to time.

For the Dicke model with strong atom-photon coupling, it is reasonable to transform to the $x$-$p$ representation in terms of real fields
\eq{Nambu7}{
x(t)=\frac{1}{\sqrt{2}}\left(a^{*}(t)+a(t)\right), \ \ \ p(t)=\frac{1}{\sqrt{2}i}\left(a^{*}(t)-a(t)\right).}
This is done via the unitary transformation for the fields
\eq{Nambu8}{
\left(x(t), p(t)\right)=\left(a^{\dagger}(t), a(t)\right) \underbrace{\frac{1}{\sqrt{2}} \left(\begin{array}{cc}  1& -i\\ 1&i \end{array}\right)}_{=V}}
and the Green's function
\eq{Nambu9}{G^R_{x-p}(t,t")=V^{\dagger}G^R_{2\times2}(t,t')V=\left(\begin{array}{cc} G^R_{xx}(t,t') & G^R_{xp}(t,t') \\
G^R_{px}(t,t') & G^R_{pp}(t,t') \end{array}\right).}
The same can be done for the advanced and Keldysh Green's functions, leading to the expressions for response and correlation functions as discussed in the main text.

\begin{widetext}
\section{Markovian dissipation vs. quenched disorder}\label{sec:MvsQ}
As anticipated in the main text, the quenched bath, resulting from the coupling to a static distribution, is fundamentally different from the Markovian bath, represented by the fast electromagnetic field outside the cavity. While the dynamics of the quenched bath is frozen on the time scales of the system, the dynamics of the Markovian bath happens on much faster time scales than those of the system. As we will see, both types of bath inherently lead to non-equilibrium dynamics of the system since the system-bath equilibration time becomes infinite. For both cases this implies a non-equilibrium fluctuation-dissipation-relation (FDR), connecting response and correlations via a non-thermal distribution function.

\subsection{Non-equilibrium fluctuation dissipation relation}\label{sec:NonEqFDR}
Correlation and response properties are not fully independent of each other but connected via fluctuation-dissipation relations, which we will briefly introduce in this part. 
In a system with multiple degrees of freedom, the response properties are encoded in the retarded (advanced) Green's function ${G^{R(A)}(t,t')}$, which is defined as
\eq{MD10}{
G^R_{ij}(t,t')=-i\theta(t-t')\langle [\ann{a}{i}(t),\cre{a}{j}(t')]\rangle,}
with the commutator ${[\cdot,\cdot]}$, the system creation and annihilation operators $\cre{a}{i}, \ann{a}{i}$ and ${G^R(t,t')=\left(G^A(t',t)\right)^{\dagger}}$. The correlation function on the other hand
\eq{MD11}{
\mathcal{C}_{ij}(t,t')=\langle\{\ann{a}{i}(t),\cre{a}{j}(t')\}\rangle=iG^K_{ij}(t,t')}
is defined via the anti-commutator ${\{\cdot,\cdot\}}$ and defines the Keldysh Green's function ${G^K(t,t')}$ \cite{Kamenev2009, KamenevBook, AltlandBook}.

The fluctuation dissipation relation states
\eq{MD12}{
G^K(\omega)=G^R(\omega)F(\omega)-F(\omega)G^A(\omega)}
and relates the response and correlations of the system via the distribution function ${F(\omega)}$. In thermal equilibrium, the distribution function is fully determined by the quantum statistics of the particles and the temperature $T$ according to
\eq{MD13}{
F_{ij}(\omega)=\delta_{ij}\left(2n_{B}(\omega)+1\right),}
with the Bose distribution function $n_{B}$. As a result, in equilibrium, it is sufficient to determine either response or correlation properties in order to gain information on each of these. 

\subsection{Effective system-only action}
In this part, we present a derivation of a system-only action after elimination of the bath variables via Gaussian integration. Depending on the nature of the bath, different distribution functions will be imprinted to the system. We start with the general action of the bath, which we consider to be well described by a quadratic action and in the $(\pm)$-basis
\eq{BEq1}{
S\sub{B}=
\sum_{\mu}\int_{t,t'}\left(\cre{\zeta}{+\mu}(t),\cre{\zeta}{-\mu}(t)\right)\left(\begin{array}{cc}G^{++}_{\mu}&G^{+-}_{\mu}\\G^{-+}_{\mu}& G^{--}_{\mu}\end{array}\right)^{-1}(t,t')\left(\begin{array}{c}\ann{\zeta}{+\mu}(t')\\ \ann{\zeta}{-\mu}(t')\end{array}\right),}
with the bath variables $\zeta_{\mu}$ and the bath mode index $\mu$, which will be chosen a continuous index below. The Green's functions for the uncoupled bath variables are assumed to be in equilibrium and read
\begin{eqnarray}
G^{+-}_{\mu}(t,t')& \equiv &  G^<_{\mu}(t,t')=-i \overline{n}(\omega_{\mu})\ e^{-i\omega_{\mu}(t-t')}\label{BEq2}\\
G^{-+}_{\mu}(t,t')& \equiv &  G^>_{\mu}(t,t')=-i(\overline{n}(\omega_{\mu})+1)\ e^{-i\omega_{\mu}(t-t')}\ \ \ \\
G^{++}_{\mu}(t,t')& \equiv &  G^T_{\mu}(t,t')=\theta(t-t')G^>_{\mu}+\theta(t'-t)G^<_{\mu}\ \ \ \ \ \ \ \\
G^{--}_{\mu}(t,t')& \equiv &  G^{\tilde{T}}_{\mu}(t,t')=\theta(t-t')G^<_{\mu}+\theta(t'-t)G^>_{\mu},
\end{eqnarray}
with the bath frequencies $\omega_{\mu}$ and the familiar Green's functions $G$-\textit{lesser}, $G$-\textit{greater}, the time-ordered and the anti-time ordered Green's function. The linear coupling between system and bath is
\eq{BEq6}{
S\sub{I}=\sum_{\mu}\sqrt{\gamma_{\mu}}\int_t \left(\cre{a}{+}(t),\cre{a}{-}(t)\right)\left(\begin{array}{cc}1&0\\0&-1\end{array}\right)\left(\begin{array}{c}\ann{\zeta}{+\mu}(t)\\ \ann{\zeta}{-\mu}(t)\end{array}\right)+\mbox{h.c.},}
where $\cre{a}{}, a$ are the system's creation and annihilation operators. For simplicity we consider only a single quantum state of the system, but a generalization to many states is straightforward. The partition function is of the general form
\begin{eqnarray}
  Z &=& \int {\cal D}[a,\cre{a}{},\ann{\zeta}{\mu},\cre{\zeta}{\mu}] e^{i (S_{\rm S} + S_{\rm I} + S_{\rm B})} \nonumber\\
  &=& \int {\cal D}[a,\cre{a}{}]  e^{iS_{\rm S}} \underbrace{\left \lbrace\int {\cal D}[\ann{\zeta}{\mu},\cre{\zeta}{\mu}] e^{i (S_{\rm I} + S_{\rm B})}  \right \rbrace}_{e^{iS\sub{eff}}}~,\label{BEq7}
\end{eqnarray}
where $S\sub{S}$ is the bare action of the system.
Now we integrate out the bath via completion of the square.  The contribution $ i S_{{\rm eff},\mu} $ of the
$\mu$th mode to the effective action reads
\begin{eqnarray}
  S_{{\rm eff},\mu}[a, \cre{a}{}] = \gamma_{\mu} \int_{t,t'} (\cre{a}{+}(t), - \cre{a}{-}(t) )
  \left ( \begin{array}{cc} ~G_{\mu}^{++}(t,t') & G_{\mu}^{-+}(t,t') \\ ~G_{\mu}^{+-}(t,t') & G_{\mu}^{--}(t,t') \end{array} \right)
 \left( \begin{array}{c} 
 \ann{a}{+}(t')\\
-  \ann{a}{-}(t')\end{array} \right)~.
\end{eqnarray}
The signs for the operators on the $-$ contour come from the backward integration in time. Thus the mixed terms will occur with an overall $-$ sign, while the $++$ and $--$ terms come with an overall $+$. Summing over all the modes $\mu$ we obtain the effective action of the bath for the field variables of the subsystem. We now take the continuum limit of densely lying bath modes, centered around some central frequency $\omega_0$ and with bandwidth $\vartheta$. That is, we substitute the sum over the modes with an integral in the energy $\Omega$ weighted by a (phenomenologically introduced) density of states (DOS) $\nu(\Omega)$ of the bath
\begin{eqnarray}
  &\sum_{\mu}\gamma_\mu \simeq \int_{\omega_0-\vartheta}^{\omega_0 + \vartheta} d\Omega& \gamma(\Omega) \nu(\Omega)
\end{eqnarray}
and obtain
\begin{eqnarray}
  S_{{\rm eff}}[a, \cre{a}{}] = - \int_{\omega_0-\vartheta}^{\omega_0 + \vartheta} d\Omega \gamma(\Omega) \nu(\Omega)\int_{t,\tau} (\cre{a}{+}(t), - \cre{a}{-}(t) )
  \left ( \begin{array}{cc} ~G_{\Omega}^{++}(\tau) & G_{\Omega}^{+-}(\tau) \\ ~G_{\Omega}^{-+}(\tau) & G_{\Omega}^{--}(\tau) \end{array} \right)
 \left( \begin{array}{c} 
 \ann{a}{+}(t-\tau )\\
 -\ann{a}{-}(t-\tau)\end{array} \right)~,\label{BEq10}
\end{eqnarray}
where in addition we have used the translation invariance of the bath Green's function, $G_\Omega^{\alpha\beta}(t,t') = G_\Omega^{\alpha\beta}(t - t')$ to suitably shift the integration variables. Eq.~\eqref{BEq10} is a general expression for an effective system action resulting from a coupling of the system to a bath of harmonic oscillators with a coupling that is linear in the bath operators. In the case of a strong separation of time scales, the effective action can be further simplified. Here we consider two extreme and opposite limiting cases, namely a Markov and a quenched disorder bath.
\subsection{The Markov approximation}\label{sec:markov}
The Markov approximation is appropriate when there exists a rotating frame in which the evolution of the system is slow compared to the scales of the bath, i.e. $\omega\sub{sys}\ll \omega_0,\vartheta$, such that the system is considered as being static on the typical time scale of the bath. This leads to a temporally local form of the resulting effective action. As an example, we derive the $(\pm)$-part of the effective action
\begin{eqnarray}
S\sub{eff}^{+-}&=&-\int dt\ \cre{a}{-}(t)\int d\tau \int_{\omega_0-\vartheta}^{\omega_0+\vartheta}\frac{d\Omega}{2\pi}\gamma(\Omega)\nu(\Omega)G^{+-}_{\Omega}(\tau)\ann{a}{+}(t-\tau)\nonumber\\
&\overset{\mbox{\tiny Markov}}\approx & -\gamma\nu\int dt\ \cre{a}{-}(t)\left(\int d\tau \int_{\omega_0-\vartheta}^{\omega_0+\vartheta}\frac{d\Omega}{2\pi}G^{+-}_{\Omega}(\tau)\right)\ann{a}{+}(t_{-\delta})\nonumber\\
&\overset{\mbox{\tiny Eq.~\eqref{BEq2}}}=&          i\gamma\nu \int dt\ \cre{a}{-}(t)\left(\int d\tau \int_{\omega_0-\vartheta}^{\omega_0+\vartheta}\frac{d\Omega}{2\pi}n(\Omega)e^{-i(\Omega-\omega_0)\tau}\right)\ann{a}{+}(t_{-\delta})      \nonumber\\
&\approx & 2i\kappa \bar{n}\int dt\ \cre{a}{-}(t)\ann{a}{+}(t_{-\delta}).
\end{eqnarray}
In the second line, we made use of the Markov approximation, i.e. the time evolution of the system is much slower than the one of the bath in the rotating frame, and the coupling and DOS are constant over the relevant frequency interval. In the third line, we replaced the Green's function by its definition (in the rotating frame). Finally, in the last line, we introduced the particle number $\bar{n}=\bar{n}(\omega_0)$ at the rotating frequency and the effective coupling $2\kappa=\gamma\nu$. Performing these steps for all the four contributions to the action in the $(\pm)$-basis leads to the action
\eq{BEq12}{S\sub{eff}[a,\cre{a}{}]= \int dt\ (\cre{a}{+}(t), \cre{a}{-}(t) ) \Sigma\sub{Mar}
 \left( \begin{array}{c} 
 \ann{a}{+}(t)\\
 \ann{a}{-}(t)\end{array} \right),}
which is local in time, containing the Markovian dissipative self-energies
 \eq{BEq13}{
 \Sigma\sub{Mar}=i\kappa\left ( \begin{array}{cc} 2\bar{n}+1 & -2(\bar{n}+1) \\-2\bar{n} & 2\bar{n}+1 \end{array} \right).}
 Transforming this self-energy to the Keldysh representation, we finally obtain
 \eq{BEq14}{
 \Sigma\sub{Mar}=i\kappa\left(\begin{array}{cc} 0 & 1\\-1 & 4\bar{n}+2\end{array}\right).}
 The additional contribution to the distribution function $F\sub{Mar}(\omega)$ for the Markovian case is obtained from the FDR for the self-energies
 \eq{BEq15}{
 \Sigma^K(\omega)=F(\omega)\left(\Sigma^R(\omega)-\Sigma^A(\omega)\right).}
 For the case when the system couples only to the Markovian bath or to an additional thermal bath, these contributions are infinitesimal and only those from the Markovian bath have to be taken into account, yielding
 \eq{BEq16}{
 i\kappa(4\bar{n}+2)=F(\omega)2i\kappa,}
 i.e. the distribution function 
 \eq{BEq17}{
 F(\omega)=2\bar{n}+1.}
 In this expression, the frequency dependent particle distribution $n(\omega)$ has been replaced by the relevant particle number $n(\omega_0)$ of the bath. The interpretation of this, is that the dynamics in the bath are so fast compared to the system, that the  for the full frequency regime, the system only couples to the slowest bath modes (in the rotating frame), located at $\omega=\omega_0$. This makes it impossible for the system to equilibrate with the bath and it can therefore not be described by a thermal distribution, i.e. stays out of equilibrium.
\subsection{The quenched bath}
The quenched bath is located in the opposite limit of the Markovian bath, i.e. it constitutes of a system bath coupling, such that there exists a rotating frame for which the system dynamics is much faster than the bath dynamics, i.e. ${\omega_0,\vartheta\ll\omega\sub{sys}}$. The corresponding approximation is to assume that the bath is static on the relevant time scale of the system and the resulting effective action for the system is infinite range in time. In this case, the contribution to the action for the $(+-)$-component reads
\begin{eqnarray}
S\sub{eff}^{+-}&=&-\int dt\ \cre{a}{-}(t)\int d\tau \int_{\omega_0-\vartheta}^{\omega_0+\vartheta}\frac{d\Omega}{2\pi}\gamma(\Omega)\nu(\Omega)G^{+-}_{\Omega}(\tau)\ann{a}{+}(t-\tau)\nonumber\\
&\overset{\mbox{\tiny quenched}}\approx &i\gamma\nu\int dt\int d\tau\ \cre{a}{-}(t)\left( \int_{\omega_0-\vartheta}^{\omega_0+\vartheta}\frac{d\Omega}{2\pi}n(\Omega)\right)\ann{a}{+}(t-\tau)      \nonumber\\
&=&2i\kappa \bar{N}\int\frac{d\omega}{2\pi} \cre{a}{-}(\omega)\ \delta(\omega)\ \ann{a}{+}(\omega).
\end{eqnarray}
In the second line, we inserted the definition of the Green's function and made the approximation of a slowly varying bath as well as a constant DOS and coupling, $\nu, \gamma$. In the third line, we replaced ${\gamma\nu=2\kappa}$ and inserted the average particle number of the bath $\bar{N}$. 

Repeating these steps for all contributions to the action in the $(\pm)$ basis and subsequently transforming to the Keldysh basis, we have the self-energy
\eq{BEq19}{
\Sigma\sub{Q}(\omega)=i\kappa\delta(\omega)\left(\begin{array}{cc} 0& 1\\-1 & 4\bar{N}+2\end{array}\right).}
This contribution is structurally different from the one from integrating out the Markovian bath, since it only acts at $\omega=0$. As a result, the distribution function for the system is only changed for $\omega=0$ compared to the uncoupled, bare system. And therefore
\eq{BEq20}{
F(\omega)=\left \{ \begin{array}{ll} 2\bar{N}+1 & \mbox{if $\omega=0$}\\
F\sub{bare}(\omega) &\mbox{if $\omega\neq0$}\end{array}\right .,}
where $F\sub{bare}$ is the distribution of the bare system. In contrast to the Markovian case, where we obtain a constant distribution for all frequencies and therefore higher system frequencies are strongly pronounced, the quenched bath shifts the occupation distribution to the very slowest modes of the system, therefore implying very slow dynamics on the system. This is reflected in the modified FDR and the appearance of a glassy phase, as discussed in Sec.~\ref{sec:calc}.

The picture obtained from these extreme cases of possible system bath couplings is quite transparent. For an equilibrium system, one assumes that the bath is such that for any possible frequency of the system, there exists a continuum of modes in the bath, such that thermalization of the system will happen on the whole frequency interval. In contrast, when the bath modes are located at much higher frequencies than the system, all the system modes interact the strongest with the slowest bath modes, leading to a distribution function as depicted in Eq.~\eqref{BEq17} and avoiding direct thermalization. On the other hand, for a bath that evolves on much slower time scales than the system, the picture is reversed, and only the slowest modes of the system interact with all the bath modes in an equivalent way. For the extreme case of a static bath, all the bath modes interact with the system's zero frequency mode, and the distribution function becomes the one in Eq.~\eqref{BEq20}. This is again a non-equilibrium distribution, such that the system does not directly thermalize.\end{widetext}

\section{Linear response in the Keldysh formalism}\label{sec:LinResp}
A common experimental procedure to probe a physical system is to apply a small external perturbation and measure the system's corresponding response. If the perturbation is sufficiently weak, the measured response will be linear in the generalized perturbing force. Here we review this construction in the Keldysh formalism in order to provide the background for the connection to the input-output formalism of quantum optics made in the text.

We consider a setup, where the hermitian operator $\hat{O}=\hat{O}^{\dagger}$ is measured after a perturbation of the form
\eq{AEq1}{
H\sub{per}(t)= F(t) \hat{O}}
has been switched on at $t=0$. Here, the (unknown) real valued field $F(t)\propto\Theta(t)$ is the corresponding generalized force.

The expectation value
\eq{AEq2}{
\langle \hat{O}\rangle(t)=\frac{1}{Z}\mbox{Tr} \left(\hat{\rho}(t) \ \hat{O}\right)}
is evaluated by introducing a source field $h(t)$, such that 
\eq{AEq3}{
\langle\hat{O}\rangle(t)=\frac{1}{Z} \left .\frac{\delta Z(h)}{\delta h(t)}\right|_{h=0},}
where 
\eq{AEq4}{
Z(h)=\mbox{Tr}\left( e^{-\beta H+\int dt\ h(t)\hat{O}(t)}\right).}
Expressing $Z$ in a real-time Keldysh framework, we have
\eq{AEq5}{
Z(h)=\int \mathcal{D}[\psi^{*},\psi] e^{iS_0[\psi^{*},\psi]} \ e^{i\delta S[h,\psi^{*},\psi]},}
where $S_0$ is the unperturbed action and $\{\psi, \psi^{*}\}$ are the complex fields representing the creation and annihilation operators of the system (in the $\pm$-basis). The term 
\begin{eqnarray}
\delta S[h,\psi^{*},\psi]&=&\int dt \left( h_{+}(t)O_{+}(t)[\psi^{*}_{+},\psi_{+}]\right .\nonumber\\
& &\left .-h_{-}(t)O_{-}(t)[\psi^{*}_{-},\psi_{-}]\right)\label{AEq6}
\end{eqnarray}
contains the source fields $h_{\pm}$ coupling to $O_{\pm}$ which polynomials in $\psi^{*},\psi$.
The expectation value \eqref{AEq2} transforms according to
\eq{AEq7}{
\langle \hat{O}(t)\rangle=\langle O_{+}(t)\rangle=\langle O_{-}(t)\rangle=\frac{1}{2}\langle O_{+}(t)+O_{-}(t)\rangle,}
 whereas the averages on the right always mean averages with respect to the functional integral.  In terms of functional derivatives of the partition function, we find
\begin{eqnarray}
\langle\hat{O}(t)\rangle&=&-\left .\frac{i}{2}\left(\frac{\delta}{\delta h_{+}(t)}-\frac{\delta}{\delta h_{-}(t)}\right)Z(h)\right|_{h=0}\nonumber\\
&=&-\left .\frac{i}{\sqrt{2}}\frac{\delta}{\delta h_{q}(t)}Z(h)\right|_{h=0}.\label{AEq8}
\end{eqnarray}
The second equality results from a rotation to the RAK representation and determines the time-dependent expectation value of $\hat{O}(t)$ for a system described by the action $S_0$. In order to incorporate the perturbation \eqref{AEq1}, we add the perturbation to the bare action of Eq.~\eqref{AEq5}
\eq{AEq9}{
S_0\longrightarrow S_0+\int dt \ \left( F_{+}(t)O_{+}(t)-F_{-}(t)O_{-}(t)\right).}

Now we can expand the expectation value of $\hat{O}$ to various orders in the force. The zeroth order simply is the expectation value in the absence of the perturbation:
\eq{AEq10}{
\langle\hat{O}(t)\rangle^{(0)}=-\left .\frac{i}{\sqrt{2}}\frac{\delta}{\delta h_q(t)} Z(h,F)\right|_{F=h=0}.}
The linear order term is then obtained via
\begin{eqnarray}
\langle\hat{O}(t)\rangle^{(1)}&=&\int_{-\infty}^t dt'\ F_+(t') \left( \frac{\delta}{\delta F_+(t')}\langle\hat{O}(t)\rangle\right)_{F=0}\nonumber\\
& &+F_-(t') \left( \frac{\delta}{\delta F_-(t')}\langle\hat{O}(t)\rangle\right)_{F=0},\label{AEq11}
\end{eqnarray}
which after a translation into the RAK representation reads
\begin{widetext}
\begin{eqnarray}
\langle\hat{O}(t)\rangle^{(1)}&=&\frac{1}{2}\int_{-\infty}^t dt'\ \left(F_+(t') \left( \frac{\delta}{\delta F_+(t')}\langle O_+(t)+O_-(t)\rangle\right)_{F=0}+F_-(t') \left( \frac{\delta}{\delta F_-(t')}\langle O_+(t)+O_-(t)\rangle\right)_{F=0}\right)\nonumber\\
&=&\frac{1}{2}\int dt'\ F(t')\left(\left(\frac{\delta}{\delta F_+(t')}+\frac{\delta}{\delta F_-(t')}\right)\langle O_+(t)+O_-(t)\rangle\right)_{F=0}\nonumber\\
&=&-\frac{i}{2}\int dt' F(t')\left .\left(\frac{\delta}{\delta F_+(t')}+\frac{\delta}{\delta F_-(t')}\right)\left(\frac{\delta}{\delta h_+(t)}-\frac{\delta}{\delta h_-(t)}\right)Z(h,F)\right|_{h=F=0}\nonumber\\
&=&-i\int dt' F(t')\left . \frac{\delta^2}{\delta F_{c}(t')\delta h_{q}(t)} Z(h,F)\right|_{F=h=0}=-\int dt' F(t') G^R_{OO}(t,t'),\label{AEq12}
\end{eqnarray}
\end{widetext}
where we made use of (at the point where we extract physical information) $F_+(t)=F_-(t)\equiv F(t)$, and furthermore that $t'\leq t$, such that the last equality indeed yields the retarded Green's function for the operator $O$. The integral in \eqref{AEq12} runs from minus infinity to plus infinity, whereas the retarded Green's function defines the upper bound being $t$ and the force $F(t')$ sets the lower bound to be $t_0$ since it vanishes for $t<t_0$ when the perturbation is switched on at $t=t_0$. Since the integral formally runs from minus infinity to plus infinity, we can switch to frequency space, where for the time-translational system (stationary state) we find 
\eq{AEqS1}{
\langle\hat{O}\rangle^{(1)}(\omega)=-F(\omega)G^R_{OO}(\omega).}
\subsection{Example: laser field induced polarization of cavity atoms}\label{sec:Polarization}
The polarization of an atomic two-level system can be expressed as 
\eq{AEq13}{
P(t)=\langle \mu_R\sigma^x(t)+\mu_I\sigma^y(t)\rangle,}
or after a rotation around the z-axis
\eq{QEq14}{
P(t)=\mu\langle\sigma^x(t)\rangle.}
We are interested in the response of the polarization to a perturbation of the system by a coherent monochromatic light field. Since the coupling of the light field is proportional to the polarization, the corresponding Hamiltonian reads
\eq{AEq15}{
H(t)=\Omega(t)\sigma^x,}
where $\Omega(t)=\theta(t)\mu E(t)$ is the generalized force and $E(t)$ is the electric field.
The corresponding action for this problem is then
\begin{eqnarray}
S&=&S_0+\delta S[h,\Omega,\phi]\nonumber\\
&=&S_0+\int dt \ h_q(t)\phi_{c}(t)+h_{c}(t)\phi_q(t)\nonumber\\
& &+\Omega_q(t)\phi_{c}(t)+\Omega_{c}(t)\phi_q(t),\label{AEq16}
\end{eqnarray}
where we have replaced $\sigma^x$ by the real fields $\phi$ as in Sec.~\ref{sec:ActionD}. 
Applying \eqref{AEq12}, we then find
\begin{eqnarray} 
P^{(1)}(t)&=&-i\left .\mu \int dt' \Omega(t')\frac{\delta^2}{\delta\Omega_{c}(t')\delta h_q(t)}Z(h,\Omega)\right|_{\Omega=h=0}\nonumber\\
&=&-\mu\int dt' \Omega(t')Q^R(t-t'),\label{AEq17}
\end{eqnarray}
where $Q^R(t-t')$ is the retarded atomic propagator as in the previous sections. 
Now we again switch to frequency space and use the definition of $\Omega$, such that we find \eq{AEq18}{ P^{(1)}(\omega)=\mu^2 E(\omega) Q^R(\omega),} where we have absorbed the $\theta$-function into the electric field. This equation identifies the retarded atomic Green's function
that we used in the previous section with the linear atomic susceptibility $\chi^{(1)}(\omega)$, which is commonly used in a quantum optics context.

\section{Distribution function of the photon $x$-component}\label{sec:XDist}
In this section, we derive the distribution function for the photonic $x$-component and show that it is identical to the atomic distribution function, proving that the atoms equilibrate with the photon $x$-component.

The Keldysh action describing the bare photon degrees of freedom is given by Eq.~\eqref{2Eq12} and we express this action directly in the Nambu basis, using the vector 
\eq{CEq1}{
\ann{A}{4}(\omega)=\left(\begin{array}{l} \ann{a}{c}(\omega)\\ \crea{a}{c}(-\omega)\\ \ann{a}{q}(\omega)\\ \crea{a}{q}(-\omega)\end{array}\right),}
the photonic action reads
\eq{CEq2}{
S\sub{ph}=\int_{\omega} \cre{A}{4}(\omega)D_{4\times 4}(\omega)\ann{A}{4}(\omega),}
with the inverse Green's function in Nambu representation
\eq{CEq3}{
D_{4\times 4}(\omega)=\left(\begin{array}{cc}   0_{2\times2}&  \left(\omega+i\kappa\right)\sigma^z+\omega_01_{2\times2}\\
\left(\omega-i\kappa\right)\sigma^z+\omega_01_{2\times2}& 2i\kappa1_{2\times2}\end{array}\right).}
The action \eqref{CEq2} can also be expressed in terms of real fields by performing the unitary transformation
\eq{CEq4}{
\left(\begin{array}{cc} x_{\alpha}(\omega)\\ p_{\alpha}(\omega)\end{array}\right)=\frac{1}{\sqrt{2}}\left(\begin{array}{cr} 1& 1\\ i & -i\end{array}\right)\left(\begin{array}{l} \ann{a}{\alpha}(\omega)\\ \crea{a}{\alpha}(-\omega)\end{array}\right),}
with $\alpha=c,q$. After this transformation, we express the action in terms of the real field
\eq{CEq5}{
V_4(\omega)=\left(\begin{array}{cc} x_{c}(\omega)\\ p_{c}(\omega)\\ x_{q}(\omega)\\ p_{q}(\omega)\end{array}\right),}
such that
\eq{CEq6}{
S\sub{ph}=\int_{\omega} V_4^T(-\omega)D_{x-p}(\omega)V_4(\omega),}
with the inverse Green's function
\eq{CEq7}{
D_{x-p}(\omega)=\left(\begin{array}{cccc} 0& 0& -\omega_0 &\kappa-i\omega\\
0& 0& -\kappa+i\omega& -\omega_0\\
-\omega_0 & -\kappa-i\omega & 2i\kappa &0\\
\kappa+i\omega& -\omega_0 & 0& 2i\kappa\end{array}\right).}
The action \eqref{CEq6} is quadratic in the fields $x_{\alpha}$ and $p_{\alpha}$ and we can eliminate the $p$-fields from the action via Gaussian integration. The resulting action is 
\eq{CEq8}{
S_x=\frac{1}{\omega_0}\int_{\omega}X^T(\omega)D_x(\omega)X(\omega),}
with the field
\eq{CEq9}{
X(\omega)=\left(\begin{array}{c} x_{c}(\omega)\\ x_{q}(\omega)\end{array}\right)}
and the inverse Green's function
\eq{CEq10}{
D_x(\omega)=\left(\begin{array}{cc} 0 & \left(\omega+i\kappa\right)^2-\omega_0^2\\
\left(\omega-i\kappa\right)^2-\omega_0^2 & \frac{2i\kappa\left(\kappa^2+\omega^2+\omega_0^2\right)}{\omega_0}\end{array}\right).}
The distribution function $F_x(\omega)$ for the $x$-field is obtained via the fluctuation-dissipation relation
\eq{CEq11}{
D_x^K(\omega)=F_x(\omega)\left(D_x^R(\omega)-D_x^A(\omega)\right),}
yielding
\eq{CEq12}{
F_x(\omega)=\frac{\omega^2+\kappa^2+\omega_0^2}{2\omega_0\omega}.}
This is indeed identical to the atomic distribution function, that we have computed in Sec.~\ref{sec:AtDist}, which proves that the atoms equilibrate with the photon $x$-field.

\bibliography{SpinGlass}

\end{document}